\documentclass{aastex63}
\usepackage{bm}

\received{July 17, 2020}
\revised{September 25, 2020}
\accepted{December 12, 2020} 
\submitjournal{ApJ}

\shorttitle{RLF at $z\sim6$}
\shortauthors{Liu et al.}
\usepackage{threeparttable}
\begin{document}

\title{Constraining the quasar radio-loud fraction at $z\sim6$ with deep radio observations}

\correspondingauthor{Ran Wang}
\email{rwangkiaa@pku.edu.cn}

\author[0000-0001-9321-6000]{Yuanqi Liu}
\affiliation{Kavli Institute for Astronomy and Astrophysics, Peking University
Beijing 100871, China}
\affiliation{Department of Astronomy, School of Physics, Peking University
Beijing 100871, China}

\author{Ran Wang}
\affiliation{Kavli Institute for Astronomy and Astrophysics, Peking University
Beijing 100871, China}

\author{Emmanuel Momjian} 
\affiliation{National Radio Astronomy Observatory, P.O. Box 0, Socorro, NM 87801, USA}

\author{Eduardo Banados}
\affiliation{ Max-Planck-Institut für Astronomie, Königstuhl 17, D-69117, Heidelberg, Germany}

\author{Greg Zeimann}
\affiliation{Hobby Eberly Telescope, University of Texas, Austin, TX 78712, USA}

\author{Chris J. Willott}
\affiliation{NRC Herzberg, 5071 West Saanich Road, Victoria, BC V9E 2E7, Canada}

\author{Yoshiki Matsuoka}
\affiliation{Research Center for Space and Cosmic Evolution, Ehime University, Matsuyama, Ehime 790-8577, Japan}

\author{Alain Omont}
\affiliation{Institut d’Astrophysique de Paris, Sorbonne Université, CNRS, UMR 7095, 98 bis bd Arago, 75014 Paris, France}

\author{Yali Shao}
\affiliation{Max-Planck-Institut für Radioastronomie, Auf dem Hügel 69, 53121 Bonn, Germany}

\author{Qiong Li}
\affiliation{Kavli Institute for Astronomy and Astrophysics, Peking University
Beijing 100871, China}
\affiliation{Department of Astronomy, School of Physics, Peking University
Beijing 100871, China}

\author{Jianan Li}
\affiliation{Kavli Institute for Astronomy and Astrophysics, Peking University
Beijing 100871, China}
\affiliation{Department of Astronomy, School of Physics, Peking University
Beijing 100871, China}




\begin{abstract}
We carry out a series of deep Karl G. Jansky Very Large Array (VLA) S-band observations of a sample of 21 quasars at $z\sim6$. The new observations expand the searches of radio continuum emission to the optically faint quasar population at the highest redshift with rest-frame $4400 \rm \AA$ luminosities down to $3 \times10^{11} \ L_{\odot}$. We report the detections of two new radio-loud quasars: CFHQS J2242+0334 (hereafter J2242+0334) at $z=5.88$ and CFHQS J0227$-$0605 (hereafter J0227$-$0605) at $z=6.20$, detected with 3 GHz flux densities of $87.0 \pm 6.3 \ \mu \rm Jy$ and $55.4 \pm 6.7 \ \mu \rm Jy$, respectively. Their radio \replaced{loudness}{loudnesses} are estimated to be $54.9 \pm 4.7$ and $16.5 \pm 3.2$, respectively. To better constrain the radio-loud fraction (RLF), we combine the new measurements with the archival VLA L-band data as well as available data from the literature, considering the upper limits for non-detections and \deleted{and} possible selection effects. The final derived RLF is $9.4 \pm 5.7\%$ for the optically selected quasars at $z\sim6$. We also compare the RLF to that of the quasar samples at low redshift and check the RLF in different quasar luminosity bins. The RLF for the optically faint objects is still poorly constrained due to the limited sample size. Our \replaced{result}{results} show no evidence of significant quasar RLF evolution with redshift. There is also no clear trend of RLF evolution with quasar UV/optical luminosity due to the limited sample size of optically faint objects with deep radio observations.

\end{abstract}



\section{Introduction} \label{sec:intro}

Quasars discovered at the highest \replaced{redshifts}{redshift} significantly improve our knowledge of the formation and accretion of the first generation of supermassive black holes (SMBH) close to the end of the cosmic reionization. More than 250 quasars have been discovered at $z >$ 5.7 residing in the first billion years of the Universe. \replaced{The UV luminosity}{Their absolute magnitudes} at rest-frame 1450 $\rm\AA$ \replaced{of all detected quasars ranges in}{are in the range of}$-29.3 < M_{1450} \lesssim -22$, and the central black hole \replaced{mass ranges in}{masses are} $10^7 M_{\odot} \le M_{\rm BH} \le 10^{10} M_{\odot}$ (e.g. \citealt{Fan2003, Fan2004, Fan2006, Jiang2009, Wu2015, Jiang2015, Jiang2016, Venemans2015, Venemans2015b, Banados2016, WangF2016, WangF2019, Mazzucchelli2017}). The formation of the SMBH as massive as $10^{10} M_{\odot}$ suggests rapid SMBH accretion and significant galaxy evolution within 1 Gyr after the Big Bang \citep{Wu2015}. The broad band UV to radio spectral energy distributions (SEDs) of these earliest quasars are comparable to those of the typical optically luminous quasars at low-$z$, suggesting a similar mechanism of the AGN activity (e.g. \citealt{Jiang2006, Shen2019}). Meanwhile, more optically fainter quasars are discovered from deep optical and near-IR surveys, such as \added{the} Canada–France–Hawaii Telescope Legacy Survey (CFHTLS) and Subaru High-$z$ Exploration of Low-Luminosity Quasars (SHELLQs) project \citep{Willott2009,Willott2010, Matsuoka2018,Matsuoka2018b}. With these fainter quasars, \citet{Matsuoka2018c} presented a new luminosity function with a sample including 110 quasars at $ 5.7 \le z \le 6.5$. They fitted the luminosity function with a double power-law function and found a break magnitude of $M^*_{1450} = -24.90^{+0.75}_{-0.90}$. These fainter objects represent the less luminous/massive but more common population that are formed at the earliest epoch. Their SMBH masses and AGN luminosities are comparable to those of the major quasar population discovered at low redshifts, thus, providing an ideal sample to investigate a possible redshift evolution of AGN activities.

Based on the differences of radio to optical flux density ratio, quasars can be divided into two categories, radio-loud (RL) and radio-quiet (RQ) quasars \citep{Kellermann1989}. The definition of radio loudness is $R = f_{\rm 5 \ GHz}/f_{\rm 4400 \ \AA}$, where $f_{\rm 5 \ GHz}$ and $f_{\rm 4400 \ \AA}$ are the radio and optical flux densities at rest-frame 5 GHz and $4400 \ \rm \AA$, respectively \citep{Kellermann1989}. For quasars with \added{a} similar optical luminosity, the radio \replaced{luminosity}{luminosities} could be different by more than two orders of magnitude between RL (\replaced{$R>10$}{$R\ge10$}) and RQ ($R<10$) sources \citep{Sanders1989, Elvis1994, Onoue2019}. \deleted{The definition of radio loudness is $R = f_{\rm 5 \ GHz}/f_{\rm 4400 \ \AA}$, where $f_{\rm 5 \ GHz}$ and $f_{\rm 4400 \ \AA}$ are the radio and optical flux densities at rest-frame 5 GHz and $4400 \ \rm \AA$, respectively \citep{Kellermann1989}.} In the past two decades, radio \replaced{telescope}{telescopes} such as Karl G. Jansky Very Large Array (VLA) and radio surveys such as the Faint Images of the Radio Sky at Twenty cm (FIRST; \citealt{Becker1995, White1997}) and the NRAO VLA Sky Survey (NVSS; \citealt{Condon1998}) provided us a basic view of the radio universe especially at low redshift. The most powerful radio emission is expected to be generated in radio-loud AGNs, where relativistic jets are launched \citep{Urry1995, Kellermann2016}. In radio-quiet quasars, radio emission comes from various (or a combination of) possible mechanisms: star formation, low-power jets, accretion disk winds and/or coronal disk emission \citep{Condon2013, Kellermann2016, Panessa2019}. The dominant mechanism can be investigated with radio emission morphology, spectral slope as well as multi-band correlations (e.g. FIR/radio correlation, Neupert effect; \citealt{Yun2001, Blundell2007, Panessa2019}). 

    The radio-loud fraction (RLF) is one of the key parameters to probe AGN radio activity among the quasar population, which is typically $10\%$ in optically selected samples based on large radio surveys \citep{ Kellermann1989, Kellermann2016, Ivezic2002, Hao2014}. By stacking the imaging data from FIRST, \citet{Jiang2007} showed that RLF of quasars decreases with the \replaced{redshift rising}{increasing redshift} from 0 to 5 and increases with increasing optical luminosity. \citet{Banados2015} reported the RLF at $z \sim 6$ \replaced{being}{to be} $8.1 ^{+5.0} _{-3.2} \%$ \replaced{from 65 quasars in total, by cross–matching optical Pan-STARRS1 and radio FIRST surveys.}{based on a sample of 65 quasars which are optically selected from the Sloan Digital Sky Survey (SDSS) survey and Panoramic Survey Telescope and Rapid Response System (Pan-STARRS) with radio measurements from published deep VLA observations or from the FIRST survey.} However, \replaced{the FIRST survey is not deep enough}{the depth of the FIRST survey is insufficient} to categorized the RL and RQ objects for the optically faint quasar population. At $z=6$, the FIRST $3\sigma$ detection limit of $0.39 \rm \ mJy$ only allows detection of $R=10$ sources with $M_{1450}<-26.7$. Deeper VLA observations were carried out for 34 $z \sim 6$ quasars with typical point source $3\sigma$ sensitivity of $20 \ \mu$Jy \citep{Wang2007, Wang2008, Wang2011}, which still focused on the luminous population with average $1450 \rm \AA$ magnitude of $M_{1450} = -26.6$. Up to now, there are 8 radio-loud quasars categorized at $5.5<z<6.5$ in total (e.g. \citealt{Wang2007,Banados2018}), including 3 radio-selected quasars \citep{McGreer2006, Zeimann2011, Belladitta2020}. Therefore, the sample used in previous studies is greatly biased to the most luminous objects. 

In this paper, we present new S-band (3 GHz) VLA observations of 21 $z \sim 6$ quasars. Compared to past studies of $z \sim 6$ quasar samples at radio wavelengths, these quasars are typically fainter in the optical. We also include 13 other sources that have archival deep VLA L-band (1.4 GHz) data. With a higher sensitivity, we newly categorize two radio-loud quasars and 20 radio-quiet quasars. Combining with previous work, we provide a better constraint on RLF. We describe the observations as well as data from literature in Section \ref{sec:data}. We show the results of two newly categorized radio-loud quasars and radio loudness calculations in Section \ref{sec:results}. We discuss how to constrain RLF, and provides related comparisons in Section \ref{sec:discussion}. A summary of the main results is presented in Section \ref{sec:summary}.

For all the cosmology calculation throughout this paper, we assume a $\Lambda$CDM cosmology with $\Omega_{\rm m} = 0.3$ and $\Omega_{\Lambda} = 0.7$, and $\rm H_0 = 70 \ km \ s^{-1} Mpc^{-1}$. Magnitudes in this paper are in the AB photometric system if not specifically pointed out.

\section{Data and observations} \label{sec:data}

\subsection{VLA S-band observations \label{subsec:sample}}
\deleted{Our} \added{The} VLA S-band (2-4 GHz; center frequency 3 GHz) observations of program 18A-232 cover a sample of 21 optically faint quasars at $5.5<z<6.5$, which have rest-frame $1450\ \rm\AA$ magnitudes of $M_{\rm1450}> -25.1$. These objects were detected in deep optical surveys, including the SDSS \citep{Jiang2009, Jiang2016}, CFHQS \citep{Willott2009, Willott2010, Willott2010b} and SHELLQs \citep{Matsuoka2016, Matsuoka2018}. The VLA observations were carried out between $18^{th}$ March \replaced{to}{and} $4^{th}$ June of 2018 in A-configuration with 27 antennas. The observation time for each target is about 1.5 hours, comprising scans on flux/bandpass calibrators and loops between targets and phase calibrators. The central frequency is 3 GHz, corresponding to quasar rest-frame of 21 GHz at $z = 6$. The total 2 GHz wide band is divided into 16 spectral windows. Each spectral window is further divided into 64 spectral channels. We excluded the channels that were significantly affected by radio frequency interference (RFI), \deleted{and} resulting in a usable bandwidth of $1.2 - 1.4$ GHz. We used the Common Astronomy Software Applications package (CASA, \citealt{McMullin2007}) version 5.1.2 to edit, calibrate, and image the data. The flux density scale calibration accuracy is about $3\%$. However, the flux calibrator 3C 48 have been undergoing a flare since January 2018, which may bring an extra effect \deleted{about} \added{of} $5\%$ in \deleted{the} accuracy. We imaged the continuum emission using natural weighting. The typical \added{FWHM} synthesized beam size is $0.8 \arcsec$, and the typical \added{1$\sigma$} rms noise on the final continuum image is $6.0 \ \mu \rm Jy \ beam^{-1}$.
\deleted{Among the 21 objects, two of them,} \added{With the new observations, two of the 21 objects,} CFHQS J2242+0334 (hereafter J2242+0334) and CFHQS J0227$-$0605 (hereafter J0227$-$0605), were detected at $> 3\sigma$. The results are listed in Table \ref{observations}.

\subsection{Data from literature \label{subsec:literature}}
We also collect available VLA data for other $z \sim 6$ quasars to carry out a statistical analysis of their radio activity. There are 42 quasars at $5.5<z<6.5$, and $-24.9 > M_{1450} > -29.3$ that have published deep VLA observations in L-band with A or B configuration, with typical rms of $25 \ \mu \rm Jy \ beam^{-1}$ \citep{Carilli2004, Wang2007, Wang2008, Wang2011,Wang2017, Banados2018}. 

In addition, 24 objects were observed by the program 11A-116 (PI: Zeimann) in VLA L-band (1.4 GHz) and A configuration. We reduce the data following the same procedure described in the previous section. Eleven of these sources are also included in our VLA S-band program. The average rms of these 24 sources is $36.1 \ \mu \rm Jy \ beam^{-1}$. The typical \added{FWHM} beam size for the L-band observation is $1.3 \arcsec$. \deleted{There is only} \added{Only} one quasar, J2242+0334, \added{is} detected in the L-band observations. These results are also shown in Table \ref{observations}. 
Four of these sources were also observed in \citet{Wang2011} with rms lower than $18 \ \mu \rm Jy \ beam^{-1}$. So we adopt the measurements from \citet{Wang2011} for further analysis. \deleted{There are two quasars,} \added{\citet{Banados2015} categorized two quasars,} J1609+3041 and J2053+0047, \deleted{categorized} as radio-loud based on data from FIRST survey \deleted{\citep{Banados2015}}. Deep L-band data obtained from the program 11A-116 for these two objects provides us a much better point-source rms noise of $\sim 30 \ \mu \rm Jy$. However, neither of them is detected. Based on the $3\sigma$ upper limits of the new data, we count J1609+3041 and J2053+0047 as radio-quiet \replaced{object}{objects} in the analysis throughout this paper.

For other optically selected $z \sim 6$ quasars \deleted{that do not have} \added{without} deep VLA observations but covered by FIRST or the stripe 82 VLA survey \citep{Becker1995, Hodge2011}, we adopt the flux densities or $3\sigma$ upper limits from these surveys to constrain their radio activities.
Objects with $M_{1450}>-23.5$ are excluded in the analysis below as few radio observations are available to set a meaningful constraints on their radio \replaced{loudness}{loudnesses}.




\subsection{Summary of the sample \label{subsec:sample-summary}}
In this work, we collect 236 optically selected quasars at $5.5<z<6.5$ with luminosity $M_{1450}<-23.5$ from various surveys (e.g. \citealt{Fan2003, Fan2004, Jiang2015, Matsuoka2016, Banados2016, WangF2019}). Our new VLA S-band observations reported in this paper and L-band observations of 11A-116 cover 34 sources from this optically-selected quasar sample. Radio data are available for another 121 objects. These include 36 objects with published deep VLA observations \replaced{at}{in} L-band and 85 objects with measurements only from the FIRST or Stripe 82 survey. These 155 sources have $1450\ \rm \AA$ magnitudes in the range of $-23.5>M_{1450}>-29.3$ and redshift range of $5.5<z<6.5$. 

\begin{longrotatetable}
\begin{deluxetable*}{ccccccccccc}
\footnotesize
\tablecaption{Observation results of quasars $z \sim 6$.}
\tabletypesize{ \small (13pt)}
\label{observations}  
\tabletypesize{\small}
\tablehead{
\colhead{Name} & \colhead{$z$} & \colhead{$M_{1450}$} & \colhead{$f_{\rm 3 \ GHz}$} & \colhead{$f_{\rm 1.4 \ GHz}$} & \colhead{$L_{\rm 5 \ GHz}$}  & \colhead{W1} & \colhead{$L_{\rm 4400\AA}$} & \colhead{$R$} & \colhead{RL/RQ} & \colhead{References} \\
\colhead{}  & \colhead{} & \colhead{(mag)} & \colhead{($\mu \rm Jy$)} & \colhead{($\mu \rm Jy$)} & \colhead($10^7 \ L_{\odot}$) & \colhead{(mag)} & \colhead{$(10^{11} \ L_{\odot})$} & \colhead{} & \colhead{}  & \colhead{}  \\
\colhead{} & \colhead{(1)} & \colhead{(2)} & \colhead{(3)} & \colhead{(4)} & \colhead{(5)} & \colhead{(6)} & \colhead{(7)} & \colhead{(8)} & \colhead{(9)} & \colhead{(10)}} 
\tabletypesize{\small}
\startdata
\textbf{CFHQS J2242+0334}	&	5.88	&	$-$24.17	&	\textbf{87.0 $\pm$  6.3}	&	\textbf{195.9 $\pm$  24.7}	&		28.6 $\pm$ 2.1	&	--	&	7.1	 $\pm$  	0.3	&		54.9 $\pm$ 4.7	&	RL		&	1 $/$ 1, 2	\\	
\textbf{CFHQS J0227$-$0605}	&	6.20	&	$-$24.98	&	\textbf{55.4 $\pm$  6.7}	&	7.2 $\pm$ 32.8	&		13.0 $\pm$ 1.6	&	--	&	10.8	 $\pm$  	1.6	&		16.5 $\pm$ 3.2	&	$\rm RL^3$		&	2 $/$ 1, 2	\\	
\hline
CFHQS J0050+3445	&	6.25	&	$-$26.57	&	--	&	18.7 $\pm$  22.3	&	$<$	9.0	&	16.65 $\pm$  0.07	&	71.4	 $\pm$  	4.6	&	$<$	1.7	&	RQ		&	1 $/$ 2	\\
CFHQS J0055+0146	&	6.02	&	$-$24.49	&	$-$0.8 $\pm$  6.8	&	67.4 $\pm$  37.5	&	$<$	4.5	&	--	&	7.4	 $\pm$  	0.4	&	$<$	8.2	&	RQ		&	2 $/$ 1, 2	\\
CFHQS J0102$-$0218	&	5.95	&	$-$24.26	&	14.1 $\pm$  6.2	&	2.02 $\pm$  31.9	&	$<$	4.0	&	--	&	6.4	 $\pm$  	0.4	&	$<$	8.6	&	RQ		&	2 $/$ 1, 2	\\
SDSS J0129$-$0035	&	5.78	&	$-$24.36	&	--	&	42.0 $\pm$  30.3	&	$<$	10.3	&	--	&	8.1	 $\pm$  	0.6	&	$<$	17.4	&	Unknown		&	3  $/$2	\\
CFHQS J0136+0226	&	6.21	&	$-$24.35	&	--	&	20.5 $\pm$  33.7	&	$<$	13.4	&	--	&	7.7	 $\pm$  	0.7	&	$<$	23.8	&	Unknown		&	1 $/$ 2	\\
SDSS J014837.64+060020.0	&	5.92	&	$-$27.08	&	--	&	52.1 $\pm$  80.0	&	$<$	28.7	&	16.10 $\pm$  0.06	&	110.0	 $\pm$  	6.1	&	$<$	3.6	&	RQ		&	4  $/$ 2	\\
HSC J0206$-$0255	&	6.03	&	$-$24.91	&	7.9 $\pm$  5.8	&	120 $\pm$  183	&	$<$	3.8	&	--	&	10.6	 $\pm$  	0.1	&	$<$	4.9	&	RQ		&	5 $/$ 1, 5	\\
CFHQS J0210$-$0456	&	6.44	&	$-$24.23	&	5.1 $\pm$  11.4	&	$-$79.8 $\pm$  64.8	&	$<$	8.7	&	--	&	5.9	 $\pm$  	0.6	&	$<$	19.9	&	Unknown		&	6 $/$ 1, 2	\\
CFHQS J0221$-$0802	&	6.16	&	$-$24.40	&	3.5 $\pm$  6.2	&	33.7 $\pm$  23.9	&	$<$	4.3	&	--	&	6.1	 $\pm$  	0.3	&	$<$	9.6	&	RQ		&	1 $/$ 1, 2	\\
SDSS J023930.24$-$004505.4	&	5.82	&	$-$24.50	&	5.7 $\pm$  5.0	&	46.3 $\pm$  41.3	&	$<$	3.1	&	--	&	7.6	 $\pm$  	0.5	&	$<$	5.5	&	RQ		&	3 $/$ 1, 2	\\
CFHQS J0316$-$1340	&	5.99	&	$-$24.58	&	--	&	10.9 $\pm$  27.1	&	$<$	10.0	&	--	&	9.6	 $\pm$  	0.7	&	$<$	14.2	&	Unknown		&	1$/$2	\\
HSC J0859+0022	&	6.39	&	$-$23.59	&	11.8 $\pm$  5.5	&	--	&	$<$	4.1	&	--	&	3.4	 $\pm$  	0.2	&	$<$	16.5	&	Unknown		&	7 $/$ 1 	\\
CFHQS J1059$-$0906	&	5.92	&	$-$25.53	&	--	&	$-$89.9 $\pm$  60.5	&	$<$	21.7	&	17.38 $\pm$  0.17	&	33.7	 $\pm$  	5.3	&	$<$	8.8	&	RQ		&	1$/$2	\\
HSC J1152+0055	&	6.37	&	$-$24.97	&	4.5 $\pm$  6.2	&	--	&	$<$	4.6	&	--	&	12.4	 $\pm$  	0.5	&	$<$	5.1	&	RQ		&	7 $/$ 1	\\
HSC J1201+0133	&	6.06	&	$-$23.85	&	0.4 $\pm$  5.0	&	230 $\pm$  169	&	$<$	3.3	&	--	&	3.5	 $\pm$  	0.1	&	$<$	13.0	&	Unknown		&	5 $/$ 1, 5	\\
SDSS J120737.43+063010.1	&	6.04	&	$-$26.60	&	  	&	57.1 $\pm$  33.6	&	$<$	12.6	&	16.99 $\pm$  0.13	&	49.7	 $\pm$  	6.0	&	$<$	3.4	&	RQ		&	4   $/$ 2	\\
HSC J1208$-$0200	&	6.20	&	$-$24.73	&	5.2 $\pm$  5.0	&	150 $\pm$  137	&	$<$	3.5	&	--	&	8.8	 $\pm$  	0.2	&	$<$	5.4	&	RQ		&	8 $/$ 1, 5	\\
ULAS J131911.29+095051.40	&	6.13	&	$-$27.07	&	--	&	\textbf{64 $\pm$  17}	&		8.2 $\pm$ 2.2	&	17.03 $\pm$ 0.11	&	48.9	 $\pm$  	5.0	&		2.3 $\pm$ 0.7	&	RQ		&	9 $/$ 2, 4	\\	
CFHQS J1509$-$1749	&	6.12	&	$-$26.93	&	--	&	23 $\pm$  18	&	$<$	6.9	&	--	&	54.4	 $\pm$  	2.0	&	$<$	1.7	&	RQ		&	10 $/$ 2, 4	\\
SDSS J160937.27+304147.7	&	6.14	&	$-$26.62	&	--	&	58.2 $\pm$  23.4	&	$<$	9.1	&	17.52 $\pm$  0.14	&	31.2	 $\pm$  	4.0	&	$<$	4.0	&	RQ		&	11 $/$ 2, 3	\\
SDSS J205321.77+004706.8	&	5.92	&	$-$25.47	&	--	&	13.6 $\pm$ 33.8	&	$<$	13.9	&	18.12 $\pm$  0.32	&	17.0	 $\pm$  	0.5	&	$<$	9.7	&	RQ		&	3 $/$ 2, 3	\\
CFHQS J2100$-$1715	&	6.09	&	$-$24.98	&	6.7 $\pm$  5.3	&	$-$23.6 $\pm$  65.6	&	$<$	3.6	&	--	&	8.8	 $\pm$  	0.4	&	$<$	5.5	&	RQ		&	1 $/$ 1, 2	\\
SDSS J211951.89$-$004020.1	&	5.87	&	$-$24.73	&	15.0 $\pm$  6.4	&	--	&	$<$	4.0	&	--	&	10.2	 $\pm$  	0.6	&	$<$	5.4	&	RQ		&	11$/$1	\\
SDSS J214755.41+010755.3	&	5.81	&	$-$25.00	&	--	&	$-$28 $\pm$  18	&	$<$	6.2	&	17.63 $\pm$ 0.20	&	26.0	 $\pm$  	4.8	&	$<$	3.2	&	RQ		&	3 $/$ 2, 4	\\	
HSC J2216$-$0016	&	6.10	&	$-$23.58	&	10.0 $\pm$  6.3	&	--	&	$<$	4.3	&	--	&	3.8	 $\pm$  	0.2	&	$<$	15.2	&	Unknown		&	7 $/$ 1	\\
HSC J2228+0152	&	6.08	&	$-$24.00	&	$-$2.7 $\pm$  5.0	&	23 $\pm$  112	&	$<$	3.4	&	--	&	4.5	 $\pm$  	0.1	&	$<$	10.2	&	Unknown		&	7 $/$ 1, 5	\\
CFHQS J2229+1457	&	6.15	&	$-$24.47	&	6.7 $\pm$  5.4	&	50.8 $\pm$  32.8	&	$<$	3.7	&	--	&	8.2	 $\pm$  	0.3	&	$<$	6.2	&	RQ		&	1 $/$ 1, 2	\\
HSC J2239+0207	&	6.26	&	$-$24.69	&	10.7 $\pm$  6.0	&	74 $\pm$  108	&	$<$	4.3	&	--	&	7.5	 $\pm$  	0.1	&	$<$	7.8	&	RQ		&	5 $/$ 1, 5	\\
SDSS J230735.35+003149.4	&	5.87	&	$-$24.93	&	--	&	$-$21 $\pm$  17	&	$<$	6.0	&	17.08 $\pm$ 0.13	&	43.7	 $\pm$  	5.0	&	$<$	1.8	&	RQ		&	3 $/$ 2, 4	\\	
CFHQS J2318$-$0246	&	6.05	&	$-$24.78	&	1.2 $\pm$  5.2	&	8.5 $\pm$  43.7	&	$<$	3.5	&	--	&	10.9	 $\pm$  	0.5	&	$<$	4.3	&	RQ		&	2 $/$ 1, 2	\\
CFHQS J2329$-$0403	&	5.90	&	$-$24.31	&	$-$3.7 $\pm$  4.8	&	$-$287 $\pm$  319	&	$<$	3.0	&	--	&	7.8	 $\pm$  	0.6	&	$<$	5.3	&	RQ		&	2 $/$ 1, 5	\\
SDSS J235651.58+002333.3	&	6.00	&	$-$24.92	&	$-$3.3 $\pm$  5.8	&	19.9 $\pm$  30.9	&	$<$	3.8	&	--	&	11.4	 $\pm$  	0.5	&	$<$	4.5	&	RQ		&	3 $/$ 1, 2	\\
\enddata
\tablecomments{\\
1. Sources in bold face are the two detections in the VLA S-band observations. Values in bold face refer to $>3\sigma$ detections. \\
2. Column description: (1) $z$ -- redshift; (2) $M_{1450}$ -- absolute magnitude at rest-frame $1450 \rm \AA$, it is found from discovery papers, and we unified in our cosmological model; (3) $f_{3 \ GHz}$ -- radio flux density at observe-frame 3 GHz, on-source pixel value for non-detection sources; (4) $f_{1.4 \ GHz}$ -- radio flux density at observe-frame 1.4 GHz, on-source pixel value for non-detection sources; (5) $L_{\rm 5 \ GHz}$ -- radio luminosity at rest-frame $5 \ \rm GHz$ in unit of solar luminosity $10^7 L_{\odot}$, we adopt a spectral index of -1.07 for J2242+0334, -0.75 for J0227$-$0605 and the rest of the quasars (see Section \ref{subsec:RL} for details); (6) W1 magnitude if this source is detected, from $WISE$ data catalog, Vega magnitude; (7) $L_{ 4400 \rm \AA}$ -- optical luminosity at rest-frame $4400 \rm \AA$ in unit of solar luminosity $10^{11} L_{\odot}$; (8) $R$ -- radio loudness $R=f_{\rm 5GHz}/f_{ 4400 \rm \AA}$; (9) RL/RQ -- classification of RL (radio-loud quasar), RQ (radio-quiet quasar) and Unknown (uncategorized quasar); (10) References -- in form of Discovery paper $/$ Radio observations; the number of discovery reference paper correspond to 1 \citep{Willott2010}, 2 \citep{Willott2009}, 3 \citep{Jiang2009}, 4 \citep{Jiang2015}, 5 \citep{Matsuoka2018}, 6 \citep{Willott2010b}, 7 \citep{Matsuoka2016}, 8 \citep{Matsuoka2018c}, 9 \citep{Mortlock2009},  10 \citep{Willott2007}, 11\citep{Jiang2016}; radio observations come from 1 (this work new VLA observations at $3 \ \rm GHz$), 2 (11A-116 at $1.4 \ \rm GHz$), 3 \citep{Banados2015}, 4 \citep{Wang2011}, 5 (FIRST, \citealt{Becker1995}).\\
3. The source J0227$-$0605 is classified as a radio-loud quasar here. But the radio loudness $R$ can be smaller than 10 if the radio spectra is flatter or we adopt different UV/optical templates in estimation (see Section \ref{subsec:J0227} for details). } 
\end{deluxetable*}
\end{longrotatetable}

\clearpage

\section{Results} \label{sec:results}
\deleted{We summarize the results measured from the new VLA S-band and L-band observations in Table \ref{observations}. }
\added{Table \ref{observations} lists the measurements from the new VLA S-band and L-band observations as well as archival data, which are used to calculate radio loudness. }
For the non-detections, we list the on-pixel value at the optical quasar position and the $1\sigma$ rms. We adopt the $3\sigma$ upper limits for these objects for the analysis of radio luminosity and radio loudness in Section \ref{subsec:RL}. For the two detections, the sources are unresolved. We adopt the peak surface brightness value on the map as the total flux density of the radio source.

\subsection{CFHQS J2242+0334 \label{subsec:J2242}}
\begin{figure*}
\gridline{\fig{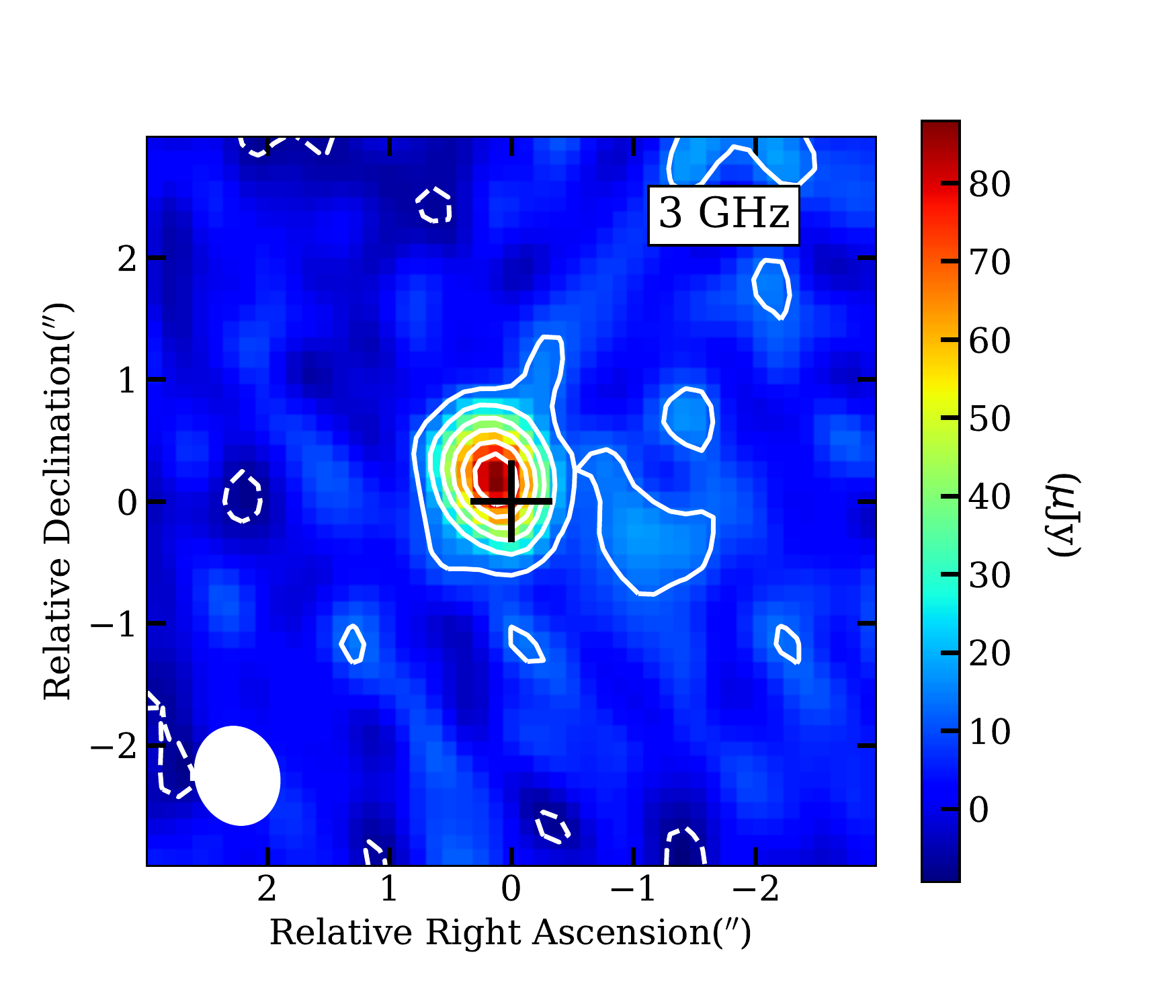}{0.52\textwidth}{(a)}
          \fig{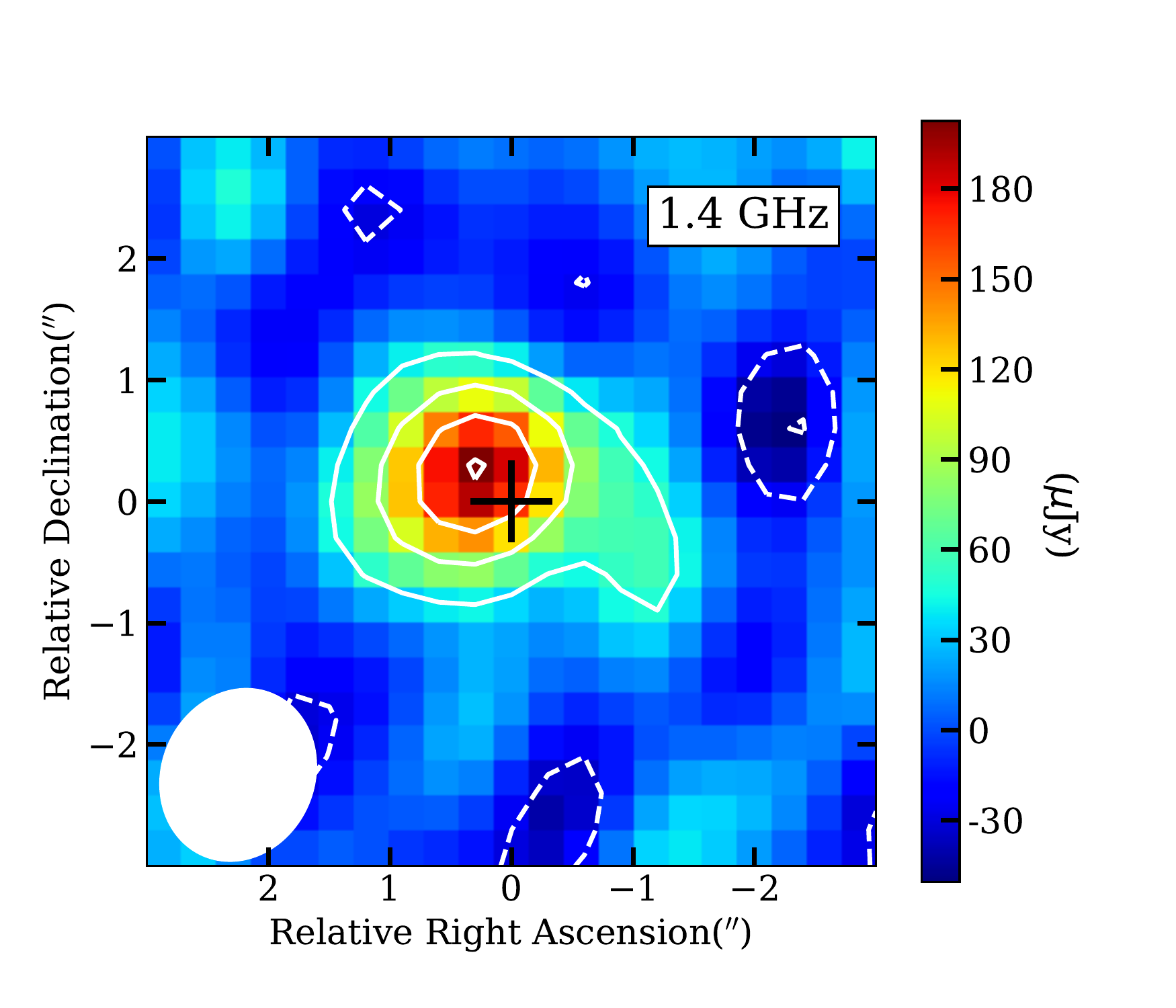}{0.52\textwidth}{(b)}
          }
\caption{J2242+0334 images in S-band (a, 3 GHz) and L-band (b, 1.4 GHz). Contours are $[-2, -1, 2, 4, 6, 8, 10, 12] \times \sigma$, where $\sigma$ is $6.3 \  \mu \rm Jy/beam$ for the 3GHz image and $24.7 \  \mu \rm Jy/beam$ for 1.4 GHz. Black crosses represent the optical \added{quasar} position (RA=22h42m37.533s, \replaced{DEC=+03d34m22.03s}{DEC=+03$\arcdeg$34$\arcmin$22.03\arcsec}). Synthesized beams are shown as ellipse at bottom left, representing the FWHM beam sizes of $0.811\arcsec \times 0.680\arcsec$ and $1.442\arcsec \times 1.247\arcsec$ in map (a) and (b), respectively.
\label{J2242}}
\end{figure*}

This is the brightest source detected in the VLA S-band observations of 21 optically faint quasars. The source is detected at $13.8 \sigma$ with a 3 GHz flux density of $f_{\rm 3 \ GHz} = 87.0 \pm 6.3 \ \mu \rm Jy$. The 3 GHz radio continuum image is shown in Figure \ref{J2242} (a) with the black cross representing the optical center of the quasar. It shows a tentative offset of $0\farcs04$ between the optical and radio center. The position uncertainty of the radio observations is estimated to be $0\farcs03$ (caused by thermal noise $\Delta \theta_{therm} \approx 0.5\theta_{\rm beam}/ \rm SNR $, where the synthesized beam is $\theta_{beam} = 0\farcs72$, \citealt{Reid2014}). And optical position measured on image with new astrometry tied to Gaia frame is \replaced{J224237.533+033422.03}{RA=22h42m37.533s, DEC=+03\arcdeg34\arcmin22.03\arcsec}, with uncertainty around $0\farcs1$. Thus, the position of the radio center is consistent with the position of the optical center.

This object is also detected at $>7\sigma$ in the L-band data with 1.4 GHz flux density of $f_{\rm 1.4 \ GHz} = 195.9 \pm 24.7 \ \mu \rm Jy$. This source is unresolved in both L- and S-bands. The measurements at 3 GHz and 1.4 GHz yield a steep power-law spectrum with spectral index $\alpha_R=-1.07^{+0.27}_{-0.25}$ $(f_{\nu} \sim \nu^{\alpha_R})$. With the optical data from \citet{Willott2010}, we calculate the radio loudness of this source to be $R_{f_{5 \ \rm GHz}/f_{4400 \rm \AA}} = 54.9 \pm 4.7$. The calculation of $f_{4400 \rm \AA}$ is described in Section \ref{subsec:RL}.


\subsection{CFHQS J0227$-$0605  \label{subsec:J0227}}
\begin{figure*}[h!]
\epsscale{0.6}
\plotone{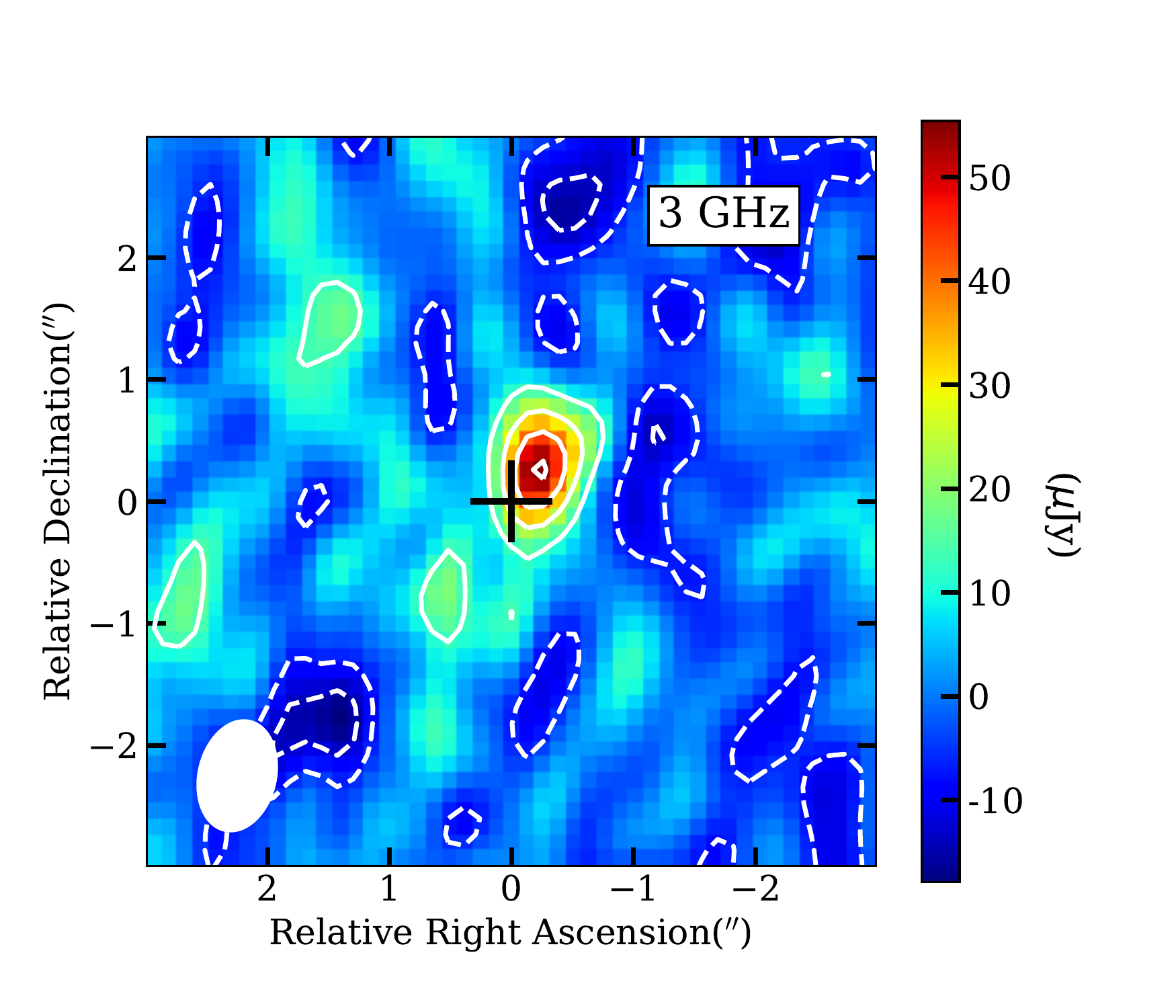}
\caption{S-band (3 GHz) image of J0227$-$0605. Contours are $[-2, -1, 2, 4, 6, 8] \times \sigma$ with rms $\sigma = 6.7 \ \mu \rm Jy \ beam^{-1}$. The black cross represents the optical quasar position \added{RA=02h27m43.320s, DEC=$-$06\arcdeg05\arcmin30.65\arcsec}. Synthesized beam with FWHM size of $0.922\arcsec \times 0.632\arcsec$ is shown as the ellipse at bottom left. 
\label{J0227}}
\end{figure*}

We detect the 3 GHz radio continuum of J0227$-$0605 at $8.3 \sigma$ with $f_{\rm 3 \ GHz} = 55.4 \pm 6.7 \ \mu \rm Jy$. The image is shown in Figure \ref{J0227}. The optical position measured with new astrometry tied to Gaia frame is \replaced{J022743.320$-$060530.65}{RA=02h27m43.320s, DEC=$-$06\arcdeg05\arcmin30.65\arcsec}, with uncertainty of $0\farcs1$ in both RA and DEC. Following the description in Section \ref{subsec:J2242}, we estimate the radio position uncertainty to be $0\farcs05$ (from synthesized beam of $0\farcs76$) for the $8.3 \sigma$ peak at 3 GHz. As shown in Figure \ref{J0227}, the 3 GHz radio peak is $0\farcs27$ away, to the northwest of the optical position. This tentative offset is slightly larger than the uncertainties of both the radio and optical positions, which should be checked with image at better spatial resolution, e.g., using the VLBA.

The source is not detected in L-band and we estimate the $3\sigma$ upper limit for the 1.4 GHz continuum flux density to be $100 \ \mu \rm Jy$. These constrain the radio spectral index to be $\alpha_R \ge -0.75$. The radio loudness of this \replaced{objects}{object} based on the 3 GHz measurement and the optical data from \citet{Willott2009} is $R = 16.5 \pm 3.2$, assuming $\alpha_R = -0.75$. The \replaced{estimation}{estimate} of $f_{4400\rm \AA}$ can be found in Section \ref{subsec:RL}. We need to point out that the radio loudness can be regarded as \replaced{a}{an} upper limit as we adopt the lower limit of radio spectral index. So we cannot rule out the possibility that J0227$-$0605 is radio-quiet if its radio spectra is flat. It requires deeper observations at 1.4 GHz or lower frequency for \added{a} better estimation. Counting J0227$-$0605 as a radio-quiet quasar will result in a lower RLF than the values presented below, in particular for the optically faint sample. But the difference is still within the uncertainties. Thus we consider J0227$-$0605 as a radio-loud quasar in the analysis below.

\subsection{Radio Loudness} \label{subsec:RL}

We estimate the \replaced{radio-loudness}{radio loudness} parameter $R= f_{\rm 5GHz}/f_{\rm 4400\AA}$ for all the sources \citep{Kellermann1989,Sikora2007}. The radio flux densities at rest-frame 5 GHz are obtained from the observed 3 GHz and/or 1.4 GHz flux densities. Only J2242+0334 is detected at both frequencies, thus we adopt its own spectral index of $\alpha_{R}=-1.07$. For J0227$-$0605, we adopt the lower limit of the radio spectral index of $\alpha_{R}=-0.75$. Note that the radio loudness value could be lower if a flatter radio spectrum is assumed. For other sources with only one detection and/or upper limit (\deleted{we adopt} \added{adopting} S-band prior to L-band detection limit for better sensitivity), we assume a power-law \deleted{spectral energy distribution} \added{spectrum} with a steep spectral index of $\alpha_{R} = -0.75$. This is widely used for quasars at $z \sim 6$ \citep{Wang2007, Banados2015}, and is consistent with results from VLBI observations \citep{Frey2011, Momjian2008, Momjian2018}.  

The optical rest-frame $\rm 4400 \ \AA$ for quasars at $z\sim6$ corresponds to an observing wavelength of $3 \ \mu m$, which is preferably obtained from near-IR and mid-IR observations. For the sources with new radio observations presented in this work, there are nine relatively luminous quasars that are detected in the $Wide$-$Field$ $Infrared \ Survey$ ($WISE$, \citealt{Wright2010}). We adopt the $W1 \ (3.5 \ \mu m)$ magnitudes for the calculation. However, other sources, especially those observed in S-band, are not covered or too faint to be detected in IR surveys, including $WISE$, $2MASS$ and $IRAC$. We collected their $M_{1450}$ (obtained from $J$-band data), $z$- and $y$-band (if have) magnitudes, which are provided in the discovery paper. We adopt the SED model in \citet{Richards2006}, fitting to the W1, or $M_{1450}$ together with $z$-/$y$-band photometric data, to obtain their optical flux densities at rest-frame $\rm4400 \ \AA$. Note that for objects with no measurements close to $3 \ \mu m$, the $\rm 4400 \ \AA$ flux densities have larger uncertainties due to the scatter of quasar UV-to-optical slope (e.g. \citealt{Richards2006}), and could be significant underestimated if the UV to optical continuum is absorbed by dust \citep{Banados2015}. 
For objects with radio data from the literature, we adopt the radio loudness values from the original paper \citep{Wang2007, Wang2008, Banados2015, Banados2018}.
 
\begin{figure}[h!]
\epsscale{0.7}
\plotone{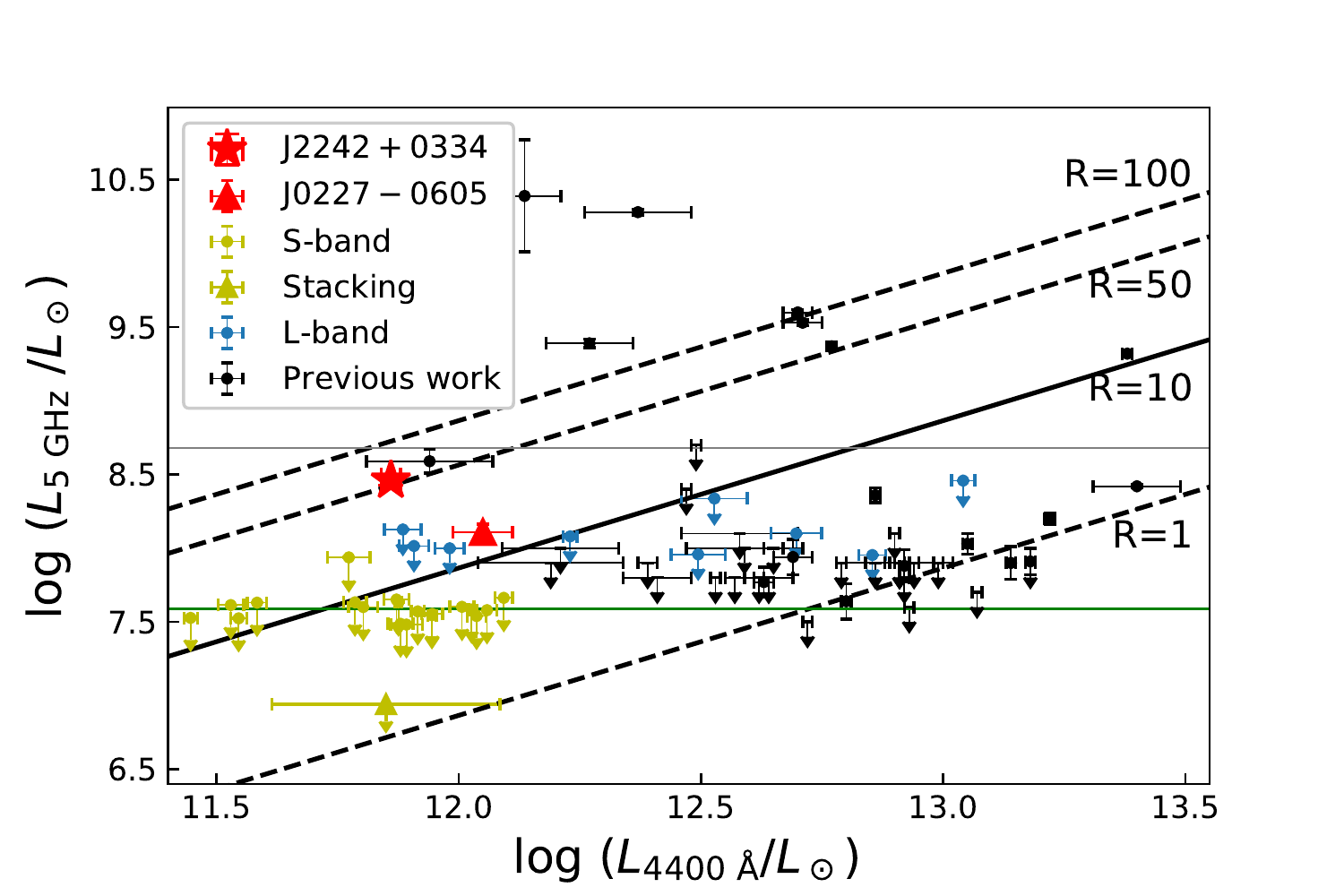}
\caption{Radio luminosity to optical luminosity plot at rest-frame 5GHz and $4400 \rm \AA$. Two VLA S-band detected quasars (red triangle and star), VLA S-band non-detections (yellow point), \replaced{VLA L-band observations}{sources observed in 11A-116} (blue point) and \replaced{former radio observations}{objects with published radio observations} (black point; \citealt{Banados2015, Banados2018, Wang2016, Belladitta2020}) are shown here. Black lines refer to different radio \replaced{loudness}{loudnesses} $R=1, 10, 50, 100$, in which the solid line of $R=10$ is the separation of radio-loud and radio-quiet sources. Horizontal green line refer to the sensitivity of VLA S-band observations, corresponds to an 5 GHz luminosity of $L_{\rm 5GHz} = 3.9 \times10^7 L_{\odot}$ at $z=6$. Grey line shows the typical $3\sigma$ detection limit of the FIRST survey, corresponds to an 5 GHz luminosity of $L_{\rm 5GHz} = 4.8 \times 10^8 L_{\odot}$ at $z=6$. Yellow triangle represents the median value stacking result of VLA S-band non-detections.  \label{rl}}
\end{figure}

We calculate the rest-frame 5 GHz radio luminosities and $\rm 4400 \ \AA$ optical luminosities with the \deleted{derived} flux densities \replaced{described}{derived} above. Figure \ref{rl} shows the radio vs optical luminosities plot for all the quasars at redshift 6 with deep radio observations. Data from previous work are plotted as black points. The depth of the FIRST survey corresponds \replaced{an}{to a} 5 GHz luminosity of $L_{\rm 5GHz} = 4.8 \times 10^8 L_{\odot}$ at $z=6$, which is shown as a grey line in Figure \ref{rl}. The typical $3\sigma$ sensitivity of new VLA S-band observations is $18 \ \mu \rm Jy$, shown as \added{a} green line, corresponding to an 5 GHz luminosity of $L_{\rm 5GHz} = 3.9 \times10^7 L_{\odot}$ at $z=6$. This is an order of magnitude deeper than the $3\sigma$ upper limit of $390 \ \mu \rm Jy$ from FIRST. The two new \replaced{radio loud}{radio-loud} quasars categorized in this work are shown with red symbols. For 
\added{the} 15 quasars that are brighter than \deleted{magnitude} $M_{1450}=-23.9$ \added{in magnitude} or \deleted{luminosity} $L_{4400\rm \ \AA} = 5.4 \times10^{11} L_{\odot}$ \added{in luminosity}, and \added{are} undetected in radio, the depths of our S-band data are sufficient to constrain their radio \replaced{loudness}{loudnesses} below the $R=10$ line. Thus we can categorize them as radio-quiet sources.

Other five sources with S-band \deleted{observation} upper limits locate above the line of $R=10$. They cannot be categorized due to their low optical luminosities ($L_{\rm 4400\AA} \le 5.4\times10^{11} \ L_{\odot}$) or noisier radio images (CFHQS J0210-0456). Furthermore, six more quasars can be newly categorized as radio-quiet sources based on the archival VLA L-band observations with $3\sigma$ sensitivity of $\sim 100 \ \mu \rm Jy$, which are shown as blue circles and located below the $R=10$ line.

The two new S-band detections, J2242+0334 and J0227$-$0605, have radio \replaced{loudness}{loudnesses} in the range of $10<R<100$. This \replaced{suggest}{suggests} that they are not as powerful as other radio-loud quasars with $R\sim1000$ (e.g. PSO J352-15, \citealt{Banados2018, Momjian2018}; CFHQS J1429+5447, \citealt{Willott2010b, Frey2011}). Such objects with moderate radio activities were sometimes called radio-intermediate quasars in the \replaced{literatures}{literature} \citep{WangTG2006, Goyal2010}.

Most of these optically faint quasars are undetected in our VLA S-band observations. In order to improve the sensitivity and better constrain the average radio emission of these objects, we constructed \added{a} stacked image for the 19 non-detections, following the procedure presented in the literature \citep{White2007,Lindroos_2014, Zwart2015, Malefahlo2020}. We cut out small stamps with sizes of $150 \times 150$ pixels centered at the quasar optical positions, and found the median value at each pixel. In the stacked image, there is no signal higher than $3\sigma$, where $\sigma = 1.3 \ \mu \rm Jy \ beam^{-1}$. We added the $3\sigma$ detection limit as a yellow triangle in Figure \ref{rl}, corresponding to $L_{\rm 5GHz} < 8.5\times 10^6 L_{\odot}$ at $z=6$ with average luminosity $L_{\rm 4400\AA} = 6.7 \times10^{11} \ L_{\odot}$. The error bar of $L_{4400\rm \AA}$ for this stacking upper limit is calculated as three times the standard deviation of the optical luminosity distribution. 
According to the 5 GHz radio luminosity to $2500\rm \AA$ absolute ultraviolet magnitude correlation of log$L_R = 0.54-0.339(M_{2500} + 25)$ ($L_R$ is the monochromatic luminosity at rest-frame 5 GHz in units of $10^{30} \ \rm ergs \ s^{-1} \ Hz^{-1}$)  from \citet{White2007} based on the SDSS and FIRST survey, an average radio luminosity of $L_{\rm 5GHz} = 3.8 \times 10^6 L_{\odot}$ ($M_{2500} =-24.7$ derived from the average luminosity $L_{\rm 4400\AA} = 6.7 \times10^{11} \ L_{\odot}$, adopting $z=6$ in the SED from \citet{Richards2006}, also applied in the following calculations) is expect for these optically faint quasars at $z\sim6$. The stacking upper limit reveals that the optically faint quasars at $z\sim 6$ are also dim in the radio, consistent with the radio-optical luminosity relation of the low-$z$ optically-selected quasars. 


\section{Discussion} \label{sec:discussion}
The deep VLA S-band data we present here largely increase the sample size of radio-observed quasars at the highest redshifts. In this section, we combine the new VLA observations with available data from literature and evaluate the RLF of these optically-selected quasars at redshift $\sim 6$. For the sample of 155 optically selected and radio-observed quasars at $5.5<z<6.5$ described in Section \ref{subsec:sample-summary}, 64 of them have radio observations that are deep enough to categorize if they are radio-loud (detections with $R\ge10$) or radio-quiet (detections or upper limits with $R<10$). \deleted{We call these 64 objects as a radio-categorized sample.} \added{These 64 objects constitute the radio categorized sample.} \replaced{For the}{The} remaining objects that are un-detected in radio with radio loudness upper limits higher than 10, \replaced{we consider them}{are named} as radio uncategorized sources. In this work, all categorized and uncategorized quasars, 155 in total, make up the all-radio sample. For comparison, we mention the 236 optically selected quasar sample at $5.5 < z < 6.5$ as the optical sample. 

For further analysis, we divide each sample into 2 luminosity bins separated at $M_{1450} = -25.5$. Sources with $-23.5 > M_{1450} > -25.5$ are classified as faint quasars, and those with $-25.5 > M_{1450} > -29.5$ are luminous quasars. We summarize these samples in Table \ref{tab:statistic}.

\begin{table}[h]
\centering
\caption{Numbers of different samples and number density based on the quasar luminosity function from  \citet{Matsuoka2018c}.}
\begin{tabular}{cccccccc}
\hline
\hline
        & -23.5$>$ $M _{1450} $ $>$-25.5 & -25.5 $>$ $M _{1450} $ & total \\
          & (faint)&  (luminous) \\ \hline
optical sample        & 74                                      & 162    &    236          \\ 
all-radio sample   & 65             & 90   & 155            \\  
radio-categorized sample   & 22        & 42    & 64               \\  
radio-loud quasars   & 2                               & 5       & 7   \\ 
\hline
number density ($\rm Gpc^{-3}$)  & $15.1 \pm 1.9$       & $2.1	 \pm 0.4$  &   \\
\hline         
\end{tabular}
\label{tab:statistic}
\end{table}


\subsection{Constraining the RLF with the radio-categorized sample
\label{subsec:direct}}

There are 5 radio-loud and 37 radio-quiet quasars categorized before this work \citep{Becker1995,Wang2007,Wang2008,Wang2011,Banados2015,Banados2018}. These quasars, together with the newly categorized 2 radio-loud and 20 radio-quiet quasars in this work, make up our radio-categorized sample.

As the radio data are collected from different programs, it is important to check whether the radio-categorized sample can represent the optical-selected quasar sample with $M_{1450} > -23.5$ and $5.5<z<6.5$. We apply t-test \citep{ttest} to the distributions of $M_{1450}$ of the radio-categorized sample and optical sample. The p-value is 0.25 ($>$0.05) indicating that there is no significant difference between these two samples.

According to Table \ref{tab:statistic}, we have 7 radio-loud objects from the radio-categorized sample of 64 quasars. This yields a fraction of radio-loud quasars to be RLF = RL/(RL+RQ) = 7/64 = $10.9 \pm 4.1\%$. The uncertainty is estimated from Poisson statistics.

If we consider the sub-sample of radio-categorized objects in two $M_{1450}$ bins separately, the fraction is $\rm RLF_{lumi}=5/42 = 11.9 \pm 5.3\%$ for the luminous sample, and $\rm RLF_{faint} = 2/22 = 9.1\pm 6.4\%$ for the faint sample.

\begin{figure*}[h]
\epsscale{0.7}
\plotone{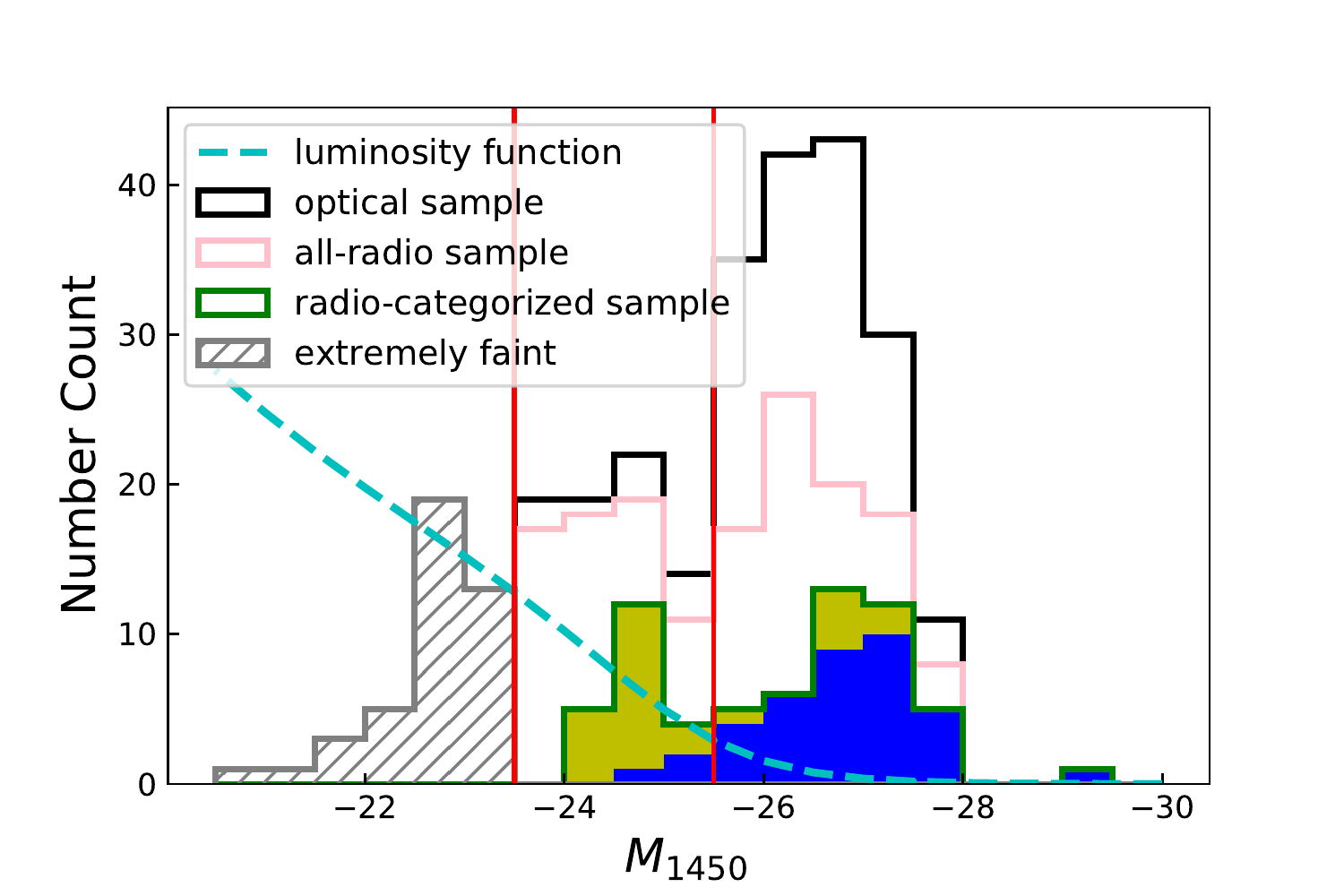}
 \caption{\replaced{Distribution of our samples in different optical luminosities.}{Distribution of M1450 for the $z\sim6$ quasar samples.} Black line represents the optical sample, while pink line represents the all-radio sample. The green line represents the radio-categorized sample, including the objects from our new S-band and L-band observations (yellow) and literature data (blue). The gray shaded area represents extremely faint quasars with $M_{1450} > -23.5$ that have insufficient radio observations and are excluded in our analysis. The cyan dashed line shows the luminosity function at redshift 6, from \citet{Matsuoka2018c}. Vertical red lines denote $M_{1450} = -23.5$ and $M_{1450} = -25.5$, which are the boundaries of faint and luminous subsamples.
\label{hist}}
\end{figure*}

One concern is that the radio-categorized sample, as well as the optical sample, is a combination of objects from optical and near-IR surveys with different detection limits. It cannot well represent the optical quasar population. As shown in Figure \ref{hist}, the number ratio between the faint and luminous radio-categorized quasars is nearly 1:2. However, the ratio of space densities of the faint and luminous optical quasars is much larger as shown below.

Here we adopt the $z=6$ quasar number density derived from the quasar luminosity function \citet{Matsuoka2018c}: 
\begin{equation}
\Phi_p(M_{1450})=\frac{\Phi^*}{10^{0.4(\alpha+1)(M_{1450}-M_{1450}^*)}+10^{0.4(\beta+1)(M_{1450}-M_{1450}^*)}} .
\end{equation}
This function is \deleted{also} plotted in Figure \ref{hist} as cyan dashed line. Integrated over the ranges of $-23.5>M_{1450}> -25.5$ and $M_{1450} >-25.5$, the number densities are $\rho_{faint} = 15.1 \pm 1.9 \ \rm Gpc^{-3}$ and $\rho_{lumi}=2.1 \pm0.4 \ \rm Gpc^{-3}$, respectively (see Table \ref{tab:statistic}). The density ratio between the faint and luminous population is about 15:2. Thus, the RLF of $10.8 \pm 4.1 \%$ derived from the whole radio-categorized sample may still have a bias to the luminous objects. Deeper observations of a much larger sample of faint objects is required to improve the statistics. Here, in order to obtain a better \replaced{estimation}{estimate} of the radio-loud faction for the $z\sim6$ quasar population with $M_{1450}<-23.5$, we weight the RLFs of the luminous and faint subsample with the quasar number densities and calculate the average RLF as :
\begin{equation}
\rm RLF_{corrected} = (\rho_{faint} \ RLF_{faint}+\rho_{lumi} \ RLF_{lumi})/ (\rho_{faint} + \rho_{lumi})= 9.4 \pm 5.7\%.     
\end{equation}
The uncertainty is propagated from poisson error of the faint and luminous subsamples.

\subsection{Constraining the RLF by all-radio sample\label{subsec:allradio}}
Here, we further consider the upper limits of $R$ of the radio-uncategorized sample of 91 quasars. As described above, these together with the radio-categorized objects constitute the all-radio sample. We repeat the t-test of $M_{1450}$ similar to Section \ref{subsec:direct} between the all-radio sample and the optical sample. The p-value = 0.09 ($>$0.05) indicates that there is no significant difference between the $M_{1450}$ distributions of these two samples. This all-radio sample can represent the optically selected quasar sample at $z\sim 6$ in the corresponding luminosity range.

\begin{figure*}[h]
\epsscale{0.7}
\plotone{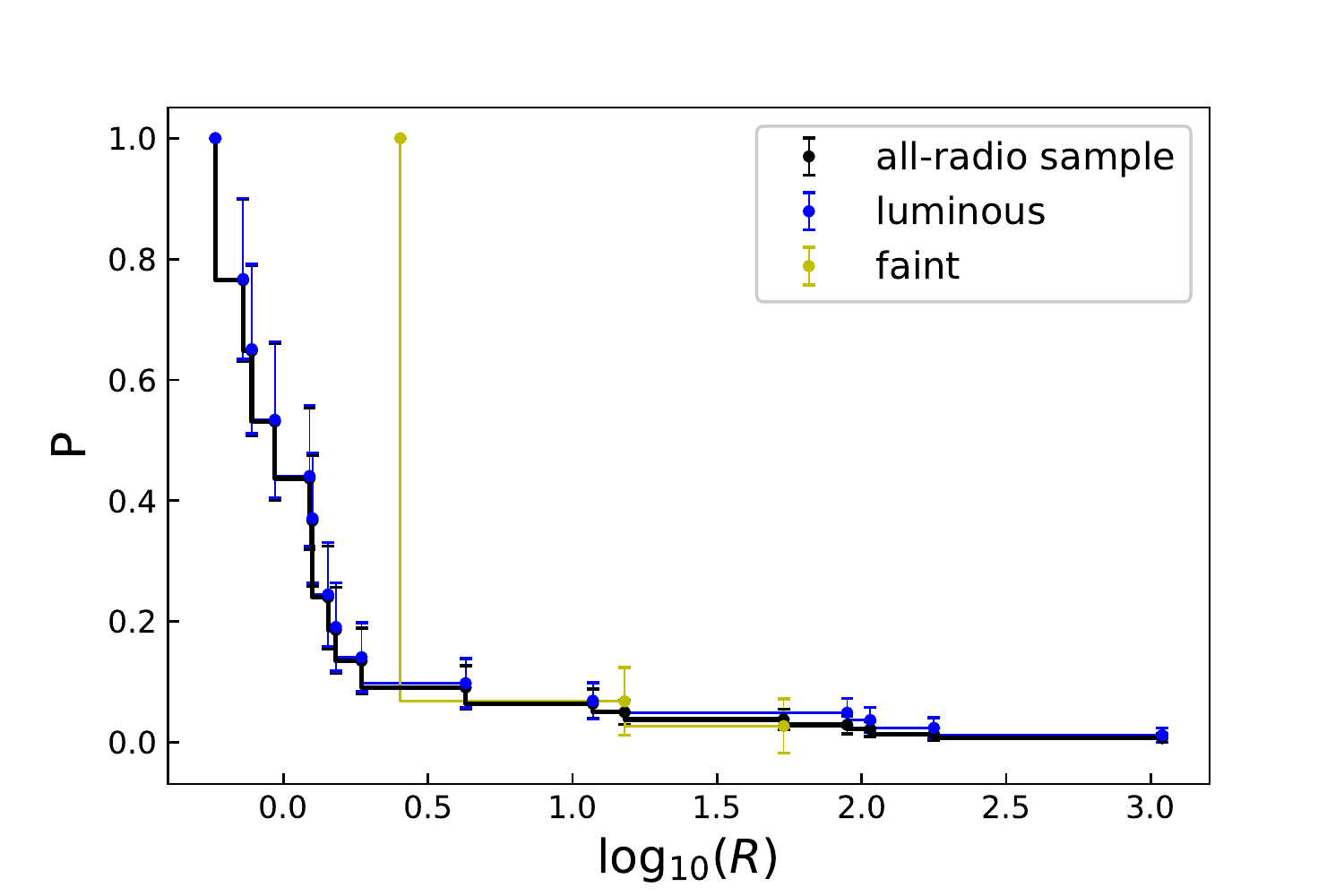}
 \caption{Survival function of radio \replaced{loudness in}{loudnesses of} all-radio sample. $P$ value \replaced{refer}{refers} to the possibility of higher than a certain radio loudness. Black line denotes the whole sample. Blue and yellow lines represent the distributions of luminous and faint sub-samples, respectively. We adopt the KM estimator to get the distribution with censored data.
 \label{KM}}
\end{figure*}

As uncategorized quasars cannot be straightly classified into the radio-loud or the radio-quiet group, we apply the Kaplan-Meier estimator (KM estimator, \citealt{KM1958,KM1}) in the analysis which can deal with censored data. The KM estimator provides a non-parameteric analysis of the radio loudness distribution, based on the detections and upper limits of the sources in the all-radio sample. The survival function is shown as the black line in Figure \ref{KM}. The y-axis label $P$ at a certain $R$ refers to the ``possibility'' that the radio loudness value is larger than $R$. Thus, the possibility $P$ at log$_{10}R>1$ refers to the possible fraction of quasars with $R > 10$, i.e., the quasar \replaced{radio loud}{radio-loud} fraction. We use the Astronomical SURVival Statistics (ASURV \added{Rev. 1.2}; \citealt{Lavalley1992}) software package to construct the distribution. Here we linearly extrapolate the $P$ value at log$_{10}R=1$ from the nearest two data points at log$_{10}R>1$ in Figure \ref{KM}. The RLF for the all radio sample can be calculated as $\rm RLF_{all-radio} = $ $P($log$_{10}R=1) = 7.3 \pm 2.2\%$ (error from the survival analysis). This result is consistent with the RLF obtained from the radio-categorized sample in Section \ref{subsec:direct}. 

\subsection{Investigating the evolution of the RLF with redshift and luminosity
\label{subsec:compare}}
The RLF for quasar samples in the local universe has been studied for more than 30 years. \citet{Kellermann1989} derived RLF around $15\% \sim 20\%$ with a sample of 114 quasars from Palomar Bright Quasar Survey (BQS) with a median redshift of 0.2. By cross matching the quasars catalog in SDSS and detections from the FIRST survey, \citet{Ivezic2002} reported an RLF of $8\pm 1\%$ with more than thousand quasars. There is no \deleted{a} clear trend of redshift evolution found with this sample (sample in redshift of $z<2.5$). A further analysis with a larger SDSS/FIRST quasar sample from z=0 to 5 suggest that the RLF of quasars decreases with increasing redshift (from 0 to 5) and decreasing optical luminosity \citep{Jiang2007, Kratzer2015}. 

\citet{Jiang2007} fitted the RLF as a function of redshift and rest-frame $2500 \ \rm \AA$ magnitude as $\rm log [RLF/(1-RLF) ] = (-0.132 \pm 0.116)+(-2.052 \pm 0.261)log(1+$$z) + (-0.183 \pm 0.025)(M_{2500}+26)$. If we adopt this function and $2500\ \rm \AA$ absolute magnitudes of $M_{2500}=-26.4$ (derived from a median $1450\ \rm \AA$ magnitudes of $-26.1$) for our $z \sim 6$ radio-categorized quasar sample, the predicted RLF at $z \sim 6$ is $1.6 \pm 2.4\%$. This is much lower than our result of $9.4 \pm 5.7\%$.  \citet{Stern2000} studied the RLF of 153 quasars at $1.75<z<2.5$ as well as 34 quasars at $4<z<4.75$, which are optically selected in a rest-frame Vega-based B band ($\lambda \approx 4400\rm \AA$) magnitude range $-26<M_{B}<-28$. The RLFs are estimated to be $13.1\pm2.7\%$ and $11.8\pm 5.5 \%$, respectively, showing no evolution in the RLFs between $z\sim2$ and $z\sim4$.
\citet{Yang2016} constrained the RLF at $z\sim 5$ from an optically luminous quasar sample with luminosity ranges in $-26.73<M_{1450}<-28.92$. They found a RLF of $7.1\%$, which also argues against a clear decrease on RLF toward the highest redshift. \citet{Banados2015} provides a constraint of RLF at $z\sim 6$ to be $8.1_{-3.2}^{+5.0} \%$ focusing on more luminous quasars. The RLF we obtained in this work is consistent with these literature values for optically selected quasars at different redshifts which do not support the redshift evolution scenario. 


\begin{table}[h]
\centering
\caption{RLF of different samples.}
\begin{tabular}{ccccccc}
\hline
\hline
        & $-23.5>M _{1450}>-25.5$ & $-25.5> M _{1450}$ & RLF&  weighted \\
          & (faint)&  (luminous) & & \\ \hline
radio-loud quasars   & 2        & 5  & & \\ 
radio-categorized sample   & $9.1 \pm 6.4\%$  & $11.9 \pm 5.3\%$  & $10.9 \pm 4.1\%$    & $9.4 \pm 5.7\%$ \\  
all-radio sample   & $8.1 \pm 3.7\%$  & $7.1 \pm 2.7\%$ & $7.3 \pm 2.2\%$ &  \\  
\citet{Jiang2007}   & $0.9\%$  & $2.2 \%$ & $1.6 \pm2.4\%$ &    \\  

\hline
\hline         
\end{tabular}
\label{tab:results}
\end{table}

We also investigate the RLF in different $M_{1450}$ bins. As shown in Table \ref{tab:results}, the differences in RLF between the faint and luminous objects are very marginal given the error bars. This is also different from the luminosity evolution scenario. e.g., based on the fitting results in \citet{Jiang2007} described above, the RLF should be $2.2 \%$ and $0.9\%$ for the luminous and faint quasar bins with $M_{1450}=-26.9$ and $M_{1450}=-24.7$, respectively. The RLF estimated from the luminous and faint bins of radio-categorized sample are $11.9 \pm 5.3\%$ and $9.1 \pm 6.4\%$, respectively. If we consider the source $J0227-0605$ as a radio-quiet quasar (discussed in Section \ref{subsec:J0227}), the RLF of the faint bin would be $4.4 \pm 4.5\%$. Here, we use the KM estimator as described in Section \ref{subsec:allradio} to calculate the RLF for the faint and luminous bins of the all-radio sample. The distribution functions of faint and luminous sub-sample are shown as yellow and blue lines in Figure \ref{KM}, which constrain the RLFs to be $8.1 \pm 3.7\%$ and $7.1 \pm 2.7\%$, respectively. We need to point out that the KM estimator may have large uncertainties when apply to small sample of objects with large percent of upper limits. In particular, the faint sub-sample contains only two radio detections, for which the KM estimator may not give a reliable estimation for the distribution of radio loudness. From both radio-categorized sample and all-radio sample, it is still difficult to draw a conclusion on how the RLF varies with quasar luminosity, under the conditions of large upper limit fraction and small sample size. Deep radio observations of a larger sample are required to reach more conclusive conclusions.

\section{Summary} \label{sec:summary}

We observe a sample of 21 optically faint quasars at \replaced{$M_{1450} < -25.1$}{$M_{1450} > -25.1$}, using VLA S-band (3 GHz) in A-configuration. Two quasars J2242+0334 and J0227$-$0605 are detected at $>3\sigma$ and categorized as radio-loud (radio-intermediate) quasars. The most powerful source in the sample, J2242+0334, is also detected with the VLA in L-band (1.4 GHz), indicating a steep spectral index.

\replaced{Our}{The} new observations provide deep radio data for the optically faint quasar population at the highest redshift, adding two radio-loud sources to current sample. In \replaced{our}{the deep} S-band observations, we categorize 14 objects as radio-quiet quasars based on the $3\sigma$ upper limits. We also reduced and analyzed archival data from the VLA program 11A-116, which observed 24 quasars at $z\sim 6$ with $1\sigma$ point source sensitivity around $30 \ \mu \rm Jy$. Furthermore, six more radio-quiet quasars have been categorized. We have constrained the RLF by this enlarged radio-categorized sample of 64 quasars to be $10.9 \pm 4.1\%$ at $z\sim 6$. 
Considering that the result may have a bias to optically luminous objects, we further calculate the RLFs in luminous and faint bins, and weight the RLFs with the quasar luminosity function. This results in a weighted-average RLF of $9.4 \pm 5.7 \%$ for the optically selected quasars with $-29.5 < M_{1450} < -23.5$ and $5.5<z<6.5$.

There are other 91 uncategorized quasars in all-radio sample. We apply the Kaplan-Meier estimator to this sample to estimate the radio loudness distribution, and derive an RLF of $7.3 \pm 2.2\%$.

The RLF obtained for the $z\sim6$ quasar sample in this work is consistent with the result from \citet{ Banados2015} which focus on the more luminous quasars at similar redshift. We \replaced{check}{investigate} the dependence of RLFs on redshift and quasar optical/UV luminosity. The RLF \added{of the $z\sim6$ quasar sample} \deleted{we} obtained \deleted{for the $z\sim6$ quasar sample} in this work is comparable to the values found with quasar samples at lower redshift \citep{Stern2000, Yang2016}. We cannot see any significant differences in RLF between samples in the optically faint and luminous bins. However, the current sample size is insufficient to well determine the RLF for the optically faint quasars at $z\sim6$. This requires further observations with a larger sample and better sensitivity.



\acknowledgments
We acknowledge the supports from the National Science Foundation of China (NSFC) grants No.11721303, 11991052, 11373008, 11533001. This work is supported by National Key Program for Science and Technology Research and Development (grant 2016YFA0400703). The National Radio Astronomy Observatory (NRAO) is a facility of the National Science Foundation operated under cooperative agreement by Associated Universities, Inc. This paper makes use of the VLA data from program 18A-232 and 11A-116.
%

\vspace{5mm}
\facilities{VLA}

\software{CASA (v5.1.2; \citealt{McMullin2007}), ASURV (\added{Rev. 1.2}; \citealt{Lavalley1992})}







\bibliography{sample63}{}

\begin{thebibliography}{}
\expandafter\ifx\csname natexlab\endcsname\relax\def\natexlab#1{#1}\fi
\providecommand{\url}[1]{\href{#1}{#1}}
\providecommand{\dodoi}[1]{doi:~\href{http://doi.org/#1}{\nolinkurl{#1}}}
\providecommand{\doeprint}[1]{\href{http://ascl.net/#1}{\nolinkurl{http://ascl.net/#1}}}
\providecommand{\doarXiv}[1]{\href{https://arxiv.org/abs/#1}{\nolinkurl{https://arxiv.org/abs/#1}}}

\bibitem[{{Ba{\~n}ados} {et~al.}(2018){Ba{\~n}ados}, {Carilli}, {Walter},
  {Momjian}, {Decarli}, {Farina}, {Mazzucchelli}, \& {Venemans}}]{Banados2018}
{Ba{\~n}ados}, E., {Carilli}, C., {Walter}, F., {et~al.} 2018, \apjl, 861, L14,
  \dodoi{10.3847/2041-8213/aac511}

\bibitem[{{Ba{\~n}ados} {et~al.}(2015){Ba{\~n}ados}, {Venemans}, {Morganson},
  {Hodge}, {Decarli}, {Walter}, {Stern}, {Schlafly}, {Farina}, {Greiner},
  {Chambers}, {Fan}, {Rix}, {Burgett}, {Draper}, {Flewelling}, {Kaiser},
  {Metcalfe}, {Morgan}, {Tonry}, \& {Wainscoat}}]{Banados2015}
{Ba{\~n}ados}, E., {Venemans}, B.~P., {Morganson}, E., {et~al.} 2015, \apj,
  804, 118, \dodoi{10.1088/0004-637X/804/2/118}

\bibitem[{Ba{\~{n}}ados {et~al.}(2016)Ba{\~{n}}ados, Venemans, Decarli, Farina,
  Mazzucchelli, Walter, Fan, Stern, Schlafly, Chambers, Rix, Jiang, McGreer,
  Simcoe, Wang, Yang, Morganson, Rosa, Greiner, Balokovi{\'{c}}, Burgett,
  Cooper, Draper, Flewelling, Hodapp, Jun, Kaiser, Kudritzki, Magnier,
  Metcalfe, Miller, Schindler, Tonry, Wainscoat, Waters, \& Yang}]{Banados2016}
Ba{\~{n}}ados, E., Venemans, B.~P., Decarli, R., {et~al.} 2016, The
  Astrophysical Journal Supplement Series, 227, 11,
  \dodoi{10.3847/0067-0049/227/1/11}

\bibitem[{{Becker} {et~al.}(1995){Becker}, {White}, \& {Helfand}}]{Becker1995}
{Becker}, R.~H., {White}, R.~L., \& {Helfand}, D.~J. 1995, \apj, 450, 559,
  \dodoi{10.1086/176166}

\bibitem[{{Belladitta} {et~al.}(2020){Belladitta}, {Moretti}, {Caccianiga},
  {Spingola}, {Severgnini}, {Della Ceca}, {Ghisellini}, {Dallacasa},
  {Sbarrato}, {Cicone}, {Cassar{\`a}}, \& {Pedani}}]{Belladitta2020}
{Belladitta}, S., {Moretti}, A., {Caccianiga}, A., {et~al.} 2020, \aap, 635,
  L7, \dodoi{10.1051/0004-6361/201937395}

\bibitem[{{Blundell} \& {Kuncic}(2007)}]{Blundell2007}
{Blundell}, K.~M., \& {Kuncic}, Z. 2007, \apjl, 668, L103,
  \dodoi{10.1086/522695}

\bibitem[{{Carilli} {et~al.}(2004){Carilli}, {Walter}, {Bertoldi}, {Menten},
  {Fan}, {Lewis}, {Strauss}, {Cox}, {Beelen}, {Omont}, \&
  {Mohan}}]{Carilli2004}
{Carilli}, C.~L., {Walter}, F., {Bertoldi}, F., {et~al.} 2004, \aj, 128, 997,
  \dodoi{10.1086/423295}

\bibitem[{{Condon} {et~al.}(1998){Condon}, {Cotton}, {Greisen}, {Yin},
  {Perley}, {Taylor}, \& {Broderick}}]{Condon1998}
{Condon}, J.~J., {Cotton}, W.~D., {Greisen}, E.~W., {et~al.} 1998, \aj, 115,
  1693, \dodoi{10.1086/300337}

\bibitem[{{Condon} {et~al.}(2013){Condon}, {Kellermann}, {Kimball},
  {Ivezi{\'c}}, \& {Perley}}]{Condon2013}
{Condon}, J.~J., {Kellermann}, K.~I., {Kimball}, A.~E., {Ivezi{\'c}}, {\v{Z}}.,
  \& {Perley}, R.~A. 2013, \apj, 768, 37, \dodoi{10.1088/0004-637X/768/1/37}

\bibitem[{{Elvis} {et~al.}(1994){Elvis}, {Wilkes}, {McDowell}, {Green},
  {Bechtold}, {Willner}, {Oey}, {Polomski}, \& {Cutri}}]{Elvis1994}
{Elvis}, M., {Wilkes}, B.~J., {McDowell}, J.~C., {et~al.} 1994, \apjs, 95, 1,
  \dodoi{10.1086/192093}

\bibitem[{{Fan} {et~al.}(2006){Fan}, {Carilli}, \& {Keating}}]{Fan2006}
{Fan}, X., {Carilli}, C.~L., \& {Keating}, B. 2006, \araa, 44, 415,
  \dodoi{10.1146/annurev.astro.44.051905.092514}

\bibitem[{{Fan} {et~al.}(2003){Fan}, {Strauss}, {Schneider}, {Becker}, {White},
  {Haiman}, {Gregg}, {Pentericci}, {Grebel}, {Narayanan}, {Loh}, {Richards},
  {Gunn}, {Lupton}, {Knapp}, {Ivezi{\'c}}, {Brandt}, {Collinge}, {Hao},
  {Harbeck}, {Prada}, {Schaye}, {Strateva}, {Zakamska}, {Anderson},
  {Brinkmann}, {Bahcall}, {Lamb}, {Okamura}, {Szalay}, \& {York}}]{Fan2003}
{Fan}, X., {Strauss}, M.~A., {Schneider}, D.~P., {et~al.} 2003, \aj, 125, 1649,
  \dodoi{10.1086/368246}

\bibitem[{{Fan} {et~al.}(2004){Fan}, {Hennawi}, {Richards}, {Strauss},
  {Schneider}, {Donley}, {Young}, {Annis}, {Lin}, {Lampeitl}, {Lupton}, {Gunn},
  {Knapp}, {Brandt}, {Anderson}, {Bahcall}, {Brinkmann}, {Brunner}, {Fukugita},
  {Szalay}, {Szokoly}, \& {York}}]{Fan2004}
{Fan}, X., {Hennawi}, J.~F., {Richards}, G.~T., {et~al.} 2004, \aj, 128, 515,
  \dodoi{10.1086/422434}

\bibitem[{{Feigelson} \& {Nelson}(1985)}]{KM1}
{Feigelson}, E.~D., \& {Nelson}, P.~I. 1985, \apj, 293, 192,
  \dodoi{10.1086/163225}

\bibitem[{{Frey} {et~al.}(2011){Frey}, {Paragi}, {Gurvits}, {Gab{\'a}nyi}, \&
  {Cseh}}]{Frey2011}
{Frey}, S., {Paragi}, Z., {Gurvits}, L.~I., {Gab{\'a}nyi}, K.~{\'E}., \&
  {Cseh}, D. 2011, \aap, 531, L5, \dodoi{10.1051/0004-6361/201117341}

\bibitem[{{Goyal} {et~al.}(2010){Goyal}, {Gopal-Krishna}, {Joshi}, {Sagar},
  {Wiita}, {Anupama}, \& {Sahu}}]{Goyal2010}
{Goyal}, A., {Gopal-Krishna}, {Joshi}, S., {et~al.} 2010, \mnras, 401, 2622,
  \dodoi{10.1111/j.1365-2966.2009.15846.x}

\bibitem[{{Hao} {et~al.}(2014){Hao}, {Sargent}, {Elvis}, {Schinnerer},
  {Zamorani}, {Ho}, {Donley}, {Civano}, {Smolcic}, {Celotti}, {Kuraszkiewicz},
  {Salvato}, {Brusa}, {Capak}, {Carilli}, {Comastri}, {Impey}, {Jahnke},
  {Koekemoer}, {Schawinski}, {Trump}, {Urry}, {Vignali}, \& {Yun}}]{Hao2014}
{Hao}, H., {Sargent}, M.~T., {Elvis}, M., {et~al.} 2014, arXiv e-prints,
  arXiv:1408.1090.
\newblock \doarXiv{1408.1090}

\bibitem[{{Hodge} {et~al.}(2011){Hodge}, {Becker}, {White}, {Richards}, \&
  {Zeimann}}]{Hodge2011}
{Hodge}, J.~A., {Becker}, R.~H., {White}, R.~L., {Richards}, G.~T., \&
  {Zeimann}, G.~R. 2011, \aj, 142, 3, \dodoi{10.1088/0004-6256/142/1/3}

\bibitem[{{Ivezi{\'c}} {et~al.}(2002){Ivezi{\'c}}, {Menou}, {Knapp}, {Strauss},
  {Lupton}, {Vand en Berk}, {Richards}, {Tremonti}, {Weinstein}, {Anderson},
  {Bahcall}, {Becker}, {Bernardi}, {Blanton}, {Eisenstein}, {Fan},
  {Finkbeiner}, {Finlator}, {Frieman}, {Gunn}, {Hall}, {Kim}, {Kinkhabwala},
  {Narayanan}, {Rockosi}, {Schlegel}, {Schneider}, {Strateva}, {SubbaRao},
  {Thakar}, {Voges}, {White}, {Yanny}, {Brinkmann}, {Doi}, {Fukugita},
  {Hennessy}, {Munn}, {Nichol}, \& {York}}]{Ivezic2002}
{Ivezi{\'c}}, {\v{Z}}., {Menou}, K., {Knapp}, G.~R., {et~al.} 2002, \aj, 124,
  2364, \dodoi{10.1086/344069}

\bibitem[{{Jiang} {et~al.}(2007){Jiang}, {Fan}, {Ivezi{\'c}}, {Richards},
  {Schneider}, {Strauss}, \& {Kelly}}]{Jiang2007}
{Jiang}, L., {Fan}, X., {Ivezi{\'c}}, {\v{Z}}., {et~al.} 2007, \apj, 656, 680,
  \dodoi{10.1086/510831}

\bibitem[{{Jiang} {et~al.}(2015){Jiang}, {McGreer}, {Fan}, {Bian}, {Cai},
  {Cl{\'e}ment}, {Wang}, \& {Fan}}]{Jiang2015}
{Jiang}, L., {McGreer}, I.~D., {Fan}, X., {et~al.} 2015, \aj, 149, 188,
  \dodoi{10.1088/0004-6256/149/6/188}

\bibitem[{{Jiang} {et~al.}(2006){Jiang}, {Fan}, {Hines}, {Shi}, {Vestergaard},
  {Bertoldi}, {Brandt}, {Carilli}, {Cox}, {Le Floc'h}, {Pentericci},
  {Richards}, {Rieke}, {Schneider}, {Strauss}, {Walter}, \&
  {Brinkmann}}]{Jiang2006}
{Jiang}, L., {Fan}, X., {Hines}, D.~C., {et~al.} 2006, \aj, 132, 2127,
  \dodoi{10.1086/508209}

\bibitem[{{Jiang} {et~al.}(2009){Jiang}, {Fan}, {Bian}, {Annis}, {Chiu},
  {Jester}, {Lin}, {Lupton}, {Richards}, {Strauss}, {Malanushenko},
  {Malanushenko}, \& {Schneider}}]{Jiang2009}
{Jiang}, L., {Fan}, X., {Bian}, F., {et~al.} 2009, \aj, 138, 305,
  \dodoi{10.1088/0004-6256/138/1/305}

\bibitem[{{Jiang} {et~al.}(2016){Jiang}, {McGreer}, {Fan}, {Strauss},
  {Ba{\~n}ados}, {Becker}, {Bian}, {Farnsworth}, {Shen}, {Wang}, {Wang},
  {Wang}, {White}, {Wu}, {Wu}, {Yang}, \& {Yang}}]{Jiang2016}
{Jiang}, L., {McGreer}, I.~D., {Fan}, X., {et~al.} 2016, \apj, 833, 222,
  \dodoi{10.3847/1538-4357/833/2/222}

\bibitem[{Kaplan \& Meier(1958)}]{KM1958}
Kaplan, E.~L., \& Meier, P. 1958, Journal of the American Statistical
  Association, 53, 457

\bibitem[{{Kellermann} {et~al.}(2016){Kellermann}, {Condon}, {Kimball},
  {Perley}, \& {Ivezi{\'c}}}]{Kellermann2016}
{Kellermann}, K.~I., {Condon}, J.~J., {Kimball}, A.~E., {Perley}, R.~A., \&
  {Ivezi{\'c}}, {\v{Z}}. 2016, \apj, 831, 168,
  \dodoi{10.3847/0004-637X/831/2/168}

\bibitem[{{Kellermann} {et~al.}(1989){Kellermann}, {Sramek}, {Schmidt},
  {Shaffer}, \& {Green}}]{Kellermann1989}
{Kellermann}, K.~I., {Sramek}, R., {Schmidt}, M., {Shaffer}, D.~B., \& {Green},
  R. 1989, \aj, 98, 1195, \dodoi{10.1086/115207}

\bibitem[{{Kratzer} \& {Richards}(2015)}]{Kratzer2015}
{Kratzer}, R.~M., \& {Richards}, G.~T. 2015, \aj, 149, 61,
  \dodoi{10.1088/0004-6256/149/2/61}

\bibitem[{{Lavalley} {et~al.}(1992){Lavalley}, {Isobe}, \&
  {Feigelson}}]{Lavalley1992}
{Lavalley}, M.~P., {Isobe}, T., \& {Feigelson}, E.~D. 1992, in \baas, Vol.~24,
  839--840

\bibitem[{Lindroos {et~al.}(2014)Lindroos, Knudsen, Vlemmings, Conway, \&
  Mart{\'\i}-Vidal}]{Lindroos_2014}
Lindroos, L., Knudsen, K.~K., Vlemmings, W., Conway, J., \& Mart{\'\i}-Vidal,
  I. 2014, Monthly Notices of the Royal Astronomical Society, 446, 3502,
  \dodoi{10.1093/mnras/stu2344}

\bibitem[{{Malefahlo} {et~al.}(2020){Malefahlo}, {Santos}, {Jarvis}, {White},
  \& {Zwart}}]{Malefahlo2020}
{Malefahlo}, E., {Santos}, M.~G., {Jarvis}, M.~J., {White}, S.~V., \& {Zwart},
  J. T.~L. 2020, \mnras, 492, 5297, \dodoi{10.1093/mnras/staa112}

\bibitem[{{Matsuoka} {et~al.}(2016){Matsuoka}, {Onoue}, {Kashikawa}, {Iwasawa},
  {Strauss}, {Nagao}, {Imanishi}, {Niida}, {Toba}, {Akiyama}, {Asami}, {Bosch},
  {Foucaud}, {Furusawa}, {Goto}, {Gunn}, {Harikane}, {Ikeda}, {Kawaguchi},
  {Kikuta}, {Komiyama}, {Lupton}, {Minezaki}, {Miyazaki}, {Morokuma},
  {Murayama}, {Nishizawa}, {Ono}, {Ouchi}, {Price}, {Sameshima}, {Silverman},
  {Sugiyama}, {Tait}, {Takada}, {Takata}, {Tanaka}, {Tang}, \&
  {Utsumi}}]{Matsuoka2016}
{Matsuoka}, Y., {Onoue}, M., {Kashikawa}, N., {et~al.} 2016, \apj, 828, 26,
  \dodoi{10.3847/0004-637X/828/1/26}

\bibitem[{{Matsuoka} {et~al.}(2018{\natexlab{a}}){Matsuoka}, {Onoue},
  {Kashikawa}, {Iwasawa}, {Strauss}, {Nagao}, {Imanishi}, {Lee}, {Akiyama},
  {Asami}, {Bosch}, {Foucaud}, {Furusawa}, {Goto}, {Gunn}, {Harikane}, {Ikeda},
  {Izumi}, {Kawaguchi}, {Kikuta}, {Kohno}, {Komiyama}, {Lupton}, {Minezaki},
  {Miyazaki}, {Morokuma}, {Murayama}, {Niida}, {Nishizawa}, {Oguri}, {Ono},
  {Ouchi}, {Price}, {Sameshima}, {Schulze}, {Shirakata}, {Silverman},
  {Sugiyama}, {Tait}, {Takada}, {Takata}, {Tanaka}, {Tang}, {Toba}, {Utsumi},
  \& {Wang}}]{Matsuoka2018}
---. 2018{\natexlab{a}}, \pasj, 70, S35, \dodoi{10.1093/pasj/psx046}

\bibitem[{{Matsuoka} {et~al.}(2018{\natexlab{b}}){Matsuoka}, {Iwasawa},
  {Onoue}, {Kashikawa}, {Strauss}, {Lee}, {Imanishi}, {Nagao}, {Akiyama},
  {Asami}, {Bosch}, {Furusawa}, {Goto}, {Gunn}, {Harikane}, {Ikeda}, {Izumi},
  {Kawaguchi}, {Kato}, {Kikuta}, {Kohno}, {Komiyama}, {Lupton}, {Minezaki},
  {Miyazaki}, {Morokuma}, {Murayama}, {Niida}, {Nishizawa}, {Oguri}, {Ono},
  {Ouchi}, {Price}, {Sameshima}, {Schulze}, {Shirakata}, {Silverman},
  {Sugiyama}, {Tait}, {Takada}, {Takata}, {Tanaka}, {Tang}, {Toba}, {Utsumi},
  {Wang}, \& {Yamashita}}]{Matsuoka2018b}
{Matsuoka}, Y., {Iwasawa}, K., {Onoue}, M., {et~al.} 2018{\natexlab{b}}, \apjs,
  237, 5, \dodoi{10.3847/1538-4365/aac724}

\bibitem[{{Matsuoka} {et~al.}(2018{\natexlab{c}}){Matsuoka}, {Strauss},
  {Kashikawa}, {Onoue}, {Iwasawa}, {Tang}, {Lee}, {Imanishi}, {Nagao},
  {Akiyama}, {Asami}, {Bosch}, {Furusawa}, {Goto}, {Gunn}, {Harikane}, {Ikeda},
  {Izumi}, {Kawaguchi}, {Kato}, {Kikuta}, {Kohno}, {Komiyama}, {Lupton},
  {Minezaki}, {Miyazaki}, {Murayama}, {Niida}, {Nishizawa}, {Noboriguchi},
  {Oguri}, {Ono}, {Ouchi}, {Price}, {Sameshima}, {Schulze}, {Shirakata},
  {Silverman}, {Sugiyama}, {Tait}, {Takada}, {Takata}, {Tanaka}, {Toba},
  {Utsumi}, {Wang}, \& {Yamashita}}]{Matsuoka2018c}
{Matsuoka}, Y., {Strauss}, M.~A., {Kashikawa}, N., {et~al.} 2018{\natexlab{c}},
  \apj, 869, 150, \dodoi{10.3847/1538-4357/aaee7a}

\bibitem[{{Mazzucchelli} {et~al.}(2017){Mazzucchelli}, {Ba{\~n}ados},
  {Venemans}, {Decarli}, {Farina}, {Walter}, {Eilers}, {Rix}, {Simcoe},
  {Stern}, {Fan}, {Schlafly}, {De Rosa}, {Hennawi}, {Chambers}, {Greiner},
  {Burgett}, {Draper}, {Kaiser}, {Kudritzki}, {Magnier}, {Metcalfe}, {Waters},
  \& {Wainscoat}}]{Mazzucchelli2017}
{Mazzucchelli}, C., {Ba{\~n}ados}, E., {Venemans}, B.~P., {et~al.} 2017, \apj,
  849, 91, \dodoi{10.3847/1538-4357/aa9185}

\bibitem[{{McGreer} {et~al.}(2006){McGreer}, {Becker}, {Helfand}, \&
  {White}}]{McGreer2006}
{McGreer}, I.~D., {Becker}, R.~H., {Helfand}, D.~J., \& {White}, R.~L. 2006,
  \apj, 652, 157, \dodoi{10.1086/507767}

\bibitem[{{McMullin} {et~al.}(2007){McMullin}, {Waters}, {Schiebel}, {Young},
  \& {Golap}}]{McMullin2007}
{McMullin}, J.~P., {Waters}, B., {Schiebel}, D., {Young}, W., \& {Golap}, K.
  2007, in Astronomical Society of the Pacific Conference Series, Vol. 376,
  Astronomical Data Analysis Software and Systems XVI, ed. R.~A. {Shaw},
  F.~{Hill}, \& D.~J. {Bell}, 127

\bibitem[{{Momjian} {et~al.}(2018){Momjian}, {Carilli}, {Ba{\~n}ados},
  {Walter}, \& {Venemans}}]{Momjian2018}
{Momjian}, E., {Carilli}, C.~L., {Ba{\~n}ados}, E., {Walter}, F., \&
  {Venemans}, B.~P. 2018, \apj, 861, 86, \dodoi{10.3847/1538-4357/aac76f}

\bibitem[{{Momjian} {et~al.}(2008){Momjian}, {Carilli}, \&
  {McGreer}}]{Momjian2008}
{Momjian}, E., {Carilli}, C.~L., \& {McGreer}, I.~D. 2008, \aj, 136, 344,
  \dodoi{10.1088/0004-6256/136/1/344}

\bibitem[{{Mortlock} {et~al.}(2009){Mortlock}, {Patel}, {Warren}, {Venemans},
  {McMahon}, {Hewett}, {Simpson}, {Sharp}, {Burningham}, {Dye}, {Ellis},
  {Gonzales-Solares}, \& {Hu{\'e}lamo}}]{Mortlock2009}
{Mortlock}, D.~J., {Patel}, M., {Warren}, S.~J., {et~al.} 2009, \aap, 505, 97,
  \dodoi{10.1051/0004-6361/200811161}

\bibitem[{{Onoue} {et~al.}(2019){Onoue}, {Kashikawa}, {Matsuoka}, {Kato},
  {Izumi}, {Nagao}, {Strauss}, {Harikane}, {Imanishi}, {Ito}, {Iwasawa},
  {Kawaguchi}, {Lee}, {Noboriguchi}, {Suh}, {Tanaka}, \& {Toba}}]{Onoue2019}
{Onoue}, M., {Kashikawa}, N., {Matsuoka}, Y., {et~al.} 2019, \apj, 880, 77,
  \dodoi{10.3847/1538-4357/ab29e9}

\bibitem[{{Panessa} {et~al.}(2019){Panessa}, {Baldi}, {Laor}, {Padovani},
  {Behar}, \& {McHardy}}]{Panessa2019}
{Panessa}, F., {Baldi}, R.~D., {Laor}, A., {et~al.} 2019, Nature Astronomy, 3,
  387, \dodoi{10.1038/s41550-019-0765-4}

\bibitem[{{Reid} \& {Honma}(2014)}]{Reid2014}
{Reid}, M.~J., \& {Honma}, M. 2014, \araa, 52, 339,
  \dodoi{10.1146/annurev-astro-081913-040006}

\bibitem[{{Richards} {et~al.}(2006){Richards}, {Lacy}, {Storrie-Lombardi},
  {Hall}, {Gallagher}, {Hines}, {Fan}, {Papovich}, {Vanden Berk}, {Trammell},
  {Schneider}, {Vestergaard}, {York}, {Jester}, {Anderson}, {Budav{\'a}ri}, \&
  {Szalay}}]{Richards2006}
{Richards}, G.~T., {Lacy}, M., {Storrie-Lombardi}, L.~J., {et~al.} 2006, \apjs,
  166, 470, \dodoi{10.1086/506525}

\bibitem[{{Sanders} {et~al.}(1989){Sanders}, {Phinney}, {Neugebauer}, {Soifer},
  \& {Matthews}}]{Sanders1989}
{Sanders}, D.~B., {Phinney}, E.~S., {Neugebauer}, G., {Soifer}, B.~T., \&
  {Matthews}, K. 1989, \apj, 347, 29, \dodoi{10.1086/168094}

\bibitem[{{Shen} {et~al.}(2019){Shen}, {Wu}, {Jiang}, {Ba{\~n}ados}, {Fan},
  {Ho}, {Riechers}, {Strauss}, {Venemans}, {Vestergaard}, {Walter}, {Wang},
  {Willott}, {Wu}, \& {Yang}}]{Shen2019}
{Shen}, Y., {Wu}, J., {Jiang}, L., {et~al.} 2019, \apj, 873, 35,
  \dodoi{10.3847/1538-4357/ab03d9}

\bibitem[{{Sikora} {et~al.}(2007){Sikora}, {Stawarz}, \& {Lasota}}]{Sikora2007}
{Sikora}, M., {Stawarz}, {\L}., \& {Lasota}, J.-P. 2007, \apj, 658, 815,
  \dodoi{10.1086/511972}

\bibitem[{{Stern} {et~al.}(2000){Stern}, {Djorgovski}, {Perley}, {de Carvalho},
  \& {Wall}}]{Stern2000}
{Stern}, D., {Djorgovski}, S.~G., {Perley}, R.~A., {de Carvalho}, R.~R., \&
  {Wall}, J.~V. 2000, \aj, 119, 1526, \dodoi{10.1086/301316}

\bibitem[{Student(1908)}]{ttest}
Student. 1908, Biometrika, 6, 1

\bibitem[{{Urry} \& {Padovani}(1995)}]{Urry1995}
{Urry}, C.~M., \& {Padovani}, P. 1995, \pasp, 107, 803, \dodoi{10.1086/133630}

\bibitem[{{Venemans} {et~al.}(2015{\natexlab{a}}){Venemans}, {Verdoes Kleijn},
  {Mwebaze}, {Valentijn}, {Ba{\~n}ados}, {Decarli}, {de Jong}, {Findlay},
  {Kuijken}, {La Barbera}, {McFarland}, {McMahon}, {Napolitano}, {Sikkema}, \&
  {Sutherland}}]{Venemans2015}
{Venemans}, B.~P., {Verdoes Kleijn}, G.~A., {Mwebaze}, J., {et~al.}
  2015{\natexlab{a}}, \mnras, 453, 2259, \dodoi{10.1093/mnras/stv1774}

\bibitem[{{Venemans} {et~al.}(2015{\natexlab{b}}){Venemans}, {Ba{\~n}ados},
  {Decarli}, {Farina}, {Walter}, {Chambers}, {Fan}, {Rix}, {Schlafly},
  {McMahon}, {Simcoe}, {Stern}, {Burgett}, {Draper}, {Flewelling}, {Hodapp},
  {Kaiser}, {Magnier}, {Metcalfe}, {Morgan}, {Price}, {Tonry}, {Waters},
  {AlSayyad}, {Banerji}, {Chen}, {Gonz{\'a}lez-Solares}, {Greiner},
  {Mazzucchelli}, {McGreer}, {Miller}, {Reed}, \& {Sullivan}}]{Venemans2015b}
{Venemans}, B.~P., {Ba{\~n}ados}, E., {Decarli}, R., {et~al.}
  2015{\natexlab{b}}, \apjl, 801, L11, \dodoi{10.1088/2041-8205/801/1/L11}

\bibitem[{{Wang} {et~al.}(2016{\natexlab{a}}){Wang}, {Wu}, {Fan}, {Yang}, {Yi},
  {Bian}, {McGreer}, {Yang}, {Ai}, {Dong}, {Zuo}, {Jiang}, {Green}, {Wang},
  {Cai}, {Wang}, \& {Yue}}]{WangF2016}
{Wang}, F., {Wu}, X.-B., {Fan}, X., {et~al.} 2016{\natexlab{a}}, \apj, 819, 24,
  \dodoi{10.3847/0004-637X/819/1/24}

\bibitem[{{Wang} {et~al.}(2019){Wang}, {Yang}, {Fan}, {Wu}, {Yue}, {Li},
  {Bian}, {Jiang}, {Ba{\~n}ados}, {Schindler}, {Findlay}, {Davies}, {Decarli},
  {Farina}, {Green}, {Hennawi}, {Huang}, {Mazzuccheli}, {McGreer}, {Venemans},
  {Walter}, {Dye}, {Lyke}, {Myers}, \& {Haze Nunez}}]{WangF2019}
{Wang}, F., {Yang}, J., {Fan}, X., {et~al.} 2019, \apj, 884, 30,
  \dodoi{10.3847/1538-4357/ab2be5}

\bibitem[{{Wang} {et~al.}(2007){Wang}, {Carilli}, {Beelen}, {Bertoldi}, {Fan},
  {Walter}, {Menten}, {Omont}, {Cox}, {Strauss}, \& {Jiang}}]{Wang2007}
{Wang}, R., {Carilli}, C.~L., {Beelen}, A., {et~al.} 2007, \aj, 134, 617,
  \dodoi{10.1086/518867}

\bibitem[{{Wang} {et~al.}(2008){Wang}, {Carilli}, {Wagg}, {Bertoldi}, {Walter},
  {Menten}, {Omont}, {Cox}, {Strauss}, {Fan}, {Jiang}, \&
  {Schneider}}]{Wang2008}
{Wang}, R., {Carilli}, C.~L., {Wagg}, J., {et~al.} 2008, \apj, 687, 848,
  \dodoi{10.1086/591076}

\bibitem[{{Wang} {et~al.}(2011){Wang}, {Wagg}, {Carilli}, {Neri}, {Walter},
  {Omont}, {Riechers}, {Bertoldi}, {Menten}, {Cox}, {Strauss}, {Fan}, \&
  {Jiang}}]{Wang2011}
{Wang}, R., {Wagg}, J., {Carilli}, C.~L., {et~al.} 2011, \aj, 142, 101,
  \dodoi{10.1088/0004-6256/142/4/101}

\bibitem[{{Wang} {et~al.}(2016{\natexlab{b}}){Wang}, {Wu}, {Neri}, {Fan},
  {Walter}, {Carilli}, {Momjian}, {Bertoldi}, {Strauss}, {Li}, {Wang},
  {Riechers}, {Jiang}, {Omont}, {Wagg}, \& {Cox}}]{Wang2016}
{Wang}, R., {Wu}, X.-B., {Neri}, R., {et~al.} 2016{\natexlab{b}}, \apj, 830,
  53, \dodoi{10.3847/0004-637X/830/1/53}

\bibitem[{{Wang} {et~al.}(2017){Wang}, {Momjian}, {Carilli}, {Wu}, {Fan},
  {Walter}, {Strauss}, {Wang}, \& {Jiang}}]{Wang2017}
{Wang}, R., {Momjian}, E., {Carilli}, C.~L., {et~al.} 2017, \apjl, 835, L20,
  \dodoi{10.3847/2041-8213/835/2/L20}

\bibitem[{{Wang} {et~al.}(2006){Wang}, {Zhou}, {Wang}, {Lu}, \&
  {Lu}}]{WangTG2006}
{Wang}, T.-G., {Zhou}, H.-Y., {Wang}, J.-X., {Lu}, Y.-J., \& {Lu}, Y. 2006,
  \apj, 645, 856, \dodoi{10.1086/504397}

\bibitem[{{White} {et~al.}(1997){White}, {Becker}, {Helfand}, \&
  {Gregg}}]{White1997}
{White}, R.~L., {Becker}, R.~H., {Helfand}, D.~J., \& {Gregg}, M.~D. 1997,
  \apj, 475, 479, \dodoi{10.1086/303564}

\bibitem[{{White} {et~al.}(2007){White}, {Helfand}, {Becker}, {Glikman}, \& {de
  Vries}}]{White2007}
{White}, R.~L., {Helfand}, D.~J., {Becker}, R.~H., {Glikman}, E., \& {de
  Vries}, W. 2007, \apj, 654, 99, \dodoi{10.1086/507700}

\bibitem[{{Willott} {et~al.}(2007){Willott}, {Delorme}, {Omont}, {Bergeron},
  {Delfosse}, {Forveille}, {Albert}, {Reyl{\'e}}, {Hill}, {Gully-Santiago},
  {Vinten}, {Crampton}, {Hutchings}, {Schade}, {Simard}, {Sawicki}, {Beelen},
  \& {Cox}}]{Willott2007}
{Willott}, C.~J., {Delorme}, P., {Omont}, A., {et~al.} 2007, \aj, 134, 2435,
  \dodoi{10.1086/522962}

\bibitem[{{Willott} {et~al.}(2009){Willott}, {Delorme}, {Reyl{\'e}}, {Albert},
  {Bergeron}, {Crampton}, {Delfosse}, {Forveille}, {Hutchings}, {McLure},
  {Omont}, \& {Schade}}]{Willott2009}
{Willott}, C.~J., {Delorme}, P., {Reyl{\'e}}, C., {et~al.} 2009, \aj, 137,
  3541, \dodoi{10.1088/0004-6256/137/3/3541}

\bibitem[{{Willott} {et~al.}(2010{\natexlab{a}}){Willott}, {Delorme},
  {Reyl{\'e}}, {Albert}, {Bergeron}, {Crampton}, {Delfosse}, {Forveille},
  {Hutchings}, {McLure}, {Omont}, \& {Schade}}]{Willott2010}
---. 2010{\natexlab{a}}, \aj, 139, 906, \dodoi{10.1088/0004-6256/139/3/906}

\bibitem[{{Willott} {et~al.}(2010{\natexlab{b}}){Willott}, {Albert},
  {Arzoumanian}, {Bergeron}, {Crampton}, {Delorme}, {Hutchings}, {Omont},
  {Reyl{\'e}}, \& {Schade}}]{Willott2010b}
{Willott}, C.~J., {Albert}, L., {Arzoumanian}, D., {et~al.} 2010{\natexlab{b}},
  \aj, 140, 546, \dodoi{10.1088/0004-6256/140/2/546}

\bibitem[{{Wright} {et~al.}(2010){Wright}, {Eisenhardt}, {Mainzer}, {Ressler},
  {Cutri}, {Jarrett}, {Kirkpatrick}, {Padgett}, {McMillan}, {Skrutskie},
  {Stanford}, {Cohen}, {Walker}, {Mather}, {Leisawitz}, {Gautier}, {McLean},
  {Benford}, {Lonsdale}, {Blain}, {Mendez}, {Irace}, {Duval}, {Liu}, {Royer},
  {Heinrichsen}, {Howard}, {Shannon}, {Kendall}, {Walsh}, {Larsen}, {Cardon},
  {Schick}, {Schwalm}, {Abid}, {Fabinsky}, {Naes}, \& {Tsai}}]{Wright2010}
{Wright}, E.~L., {Eisenhardt}, P. R.~M., {Mainzer}, A.~K., {et~al.} 2010, \aj,
  140, 1868, \dodoi{10.1088/0004-6256/140/6/1868}

\bibitem[{{Wu} {et~al.}(2015){Wu}, {Wang}, {Fan}, {Yi}, {Zuo}, {Bian}, {Jiang},
  {McGreer}, {Wang}, {Yang}, {Yang}, {Thompson}, \& {Beletsky}}]{Wu2015}
{Wu}, X.-B., {Wang}, F., {Fan}, X., {et~al.} 2015, \nat, 518, 512,
  \dodoi{10.1038/nature14241}

\bibitem[{{Yang} {et~al.}(2016){Yang}, {Wang}, {Wu}, {Fan}, {McGreer}, {Bian},
  {Yi}, {Yang}, {Ai}, {Dong}, {Zuo}, {Green}, {Jiang}, {Wang}, {Wang}, \&
  {Yue}}]{Yang2016}
{Yang}, J., {Wang}, F., {Wu}, X.-B., {et~al.} 2016, \apj, 829, 33,
  \dodoi{10.3847/0004-637X/829/1/33}

\bibitem[{{Yun} {et~al.}(2001){Yun}, {Reddy}, \& {Condon}}]{Yun2001}
{Yun}, M.~S., {Reddy}, N.~A., \& {Condon}, J.~J. 2001, \apj, 554, 803,
  \dodoi{10.1086/323145}

\bibitem[{{Zeimann} {et~al.}(2011){Zeimann}, {White}, {Becker}, {Hodge},
  {Stanford}, \& {Richards}}]{Zeimann2011}
{Zeimann}, G.~R., {White}, R.~L., {Becker}, R.~H., {et~al.} 2011, \apj, 736,
  57, \dodoi{10.1088/0004-637X/736/1/57}

\bibitem[{{Zwart} {et~al.}(2015){Zwart}, {Wall}, {Karim}, {Jackson}, {Norris},
  {Condon}, {Afonso}, {Heywood}, {Jarvis}, {Navarrete}, {Prandoni}, {Rigby},
  {Rottgering}, {Santos}, {Sargent}, {Seymour}, {Taylor}, \&
  {Vernstrom}}]{Zwart2015}
{Zwart}, J., {Wall}, J., {Karim}, A., {et~al.} 2015, in Advancing Astrophysics
  with the Square Kilometre Array (AASKA14), 172.
\newblock \doarXiv{1412.5743}

\end{thebibliography}
\bibliographystyle{aasjournal}



\end{document}